\pdfoutput=1

\documentclass[preprint]{sigplanconf}

\usepackage{amsmath}
\usepackage{comment}
\usepackage[shortlabels]{enumitem}
\usepackage[T1]{fontenc}
\usepackage[utf8x]{inputenc}
\usepackage{tikz}
\usepackage{ucs}
\usepackage{xparse}
\usepackage{microtype}
\usepackage{subfigure}

\usepackage{etoolbox} %
\newtoggle{todos}
\newtoggle{names}
\newtoggle{submit}

\newtoggle{noproofs}

\togglefalse{submit}

\iftoggle{submit}{
  \togglefalse{todos}
  \togglefalse{names}
  \toggletrue{noproofs}
}{
  \togglefalse{todos}
  \toggletrue{names}
  \toggletrue{noproofs}
}

\excludecomment{oldSec}
\iftoggle{noproofs}{
\excludecomment{optionalproof}
\excludecomment{optionallemma}
}{
\newenvironment{optionalproof}{\begin{proof}}{\end{proof}}
\newenvironment{optionallemma}{\begin{lemma}}{\end{lemma}}
}

\usepackage[scaled=0.88]{beramono} %
\usepackage{color}
\usepackage{amssymb}
\usepackage{array}
\usepackage{stmaryrd}
\usepackage{mathpartir}
\usepackage[unicode=true]{hyperref}
\usepackage{bookmark}

\iftoggle{todos}{%
\newcommand{\mynote}[2]{
  \textcolor{red}
  {
  {\bfseries\sffamily\scriptsize#1}
  {\small$\blacktriangleright$\textsf{\emph{#2}}$\blacktriangleleft$}}}
}{%
\newcommand{\mynote}[2]{}
}

\newcommand{\todo}{\mynote}

\newcommand{\pg}{\todo{PG}}

\newcommand{\yc}{\todo{YC}}

\DeclareMathAlphabet{\mathcal}{OT1}{pzc}{m}{it}

\newcommand{\Keyword}[1]{\textbf{#1}}
\newcommand{\Term}[1]{\mathit{#1}}

\NewDocumentCommand{\List}{sm}{\Parens#1{\overline{#2}}}

\NewDocumentCommand{\Index}{sm}{\IfValueTF{#2}{\IfBooleanTF{#1}{^}{_}{#2}}{}}

\NewDocumentCommand{\DIFF}{o}{\ominus\Index{#1}}
\NewDocumentCommand{\UPDATE}{o}{\oplus\Index{#1}}
\let\APPLY\UPDATE

\NewDocumentCommand{\Diff}{somm}{\Parens#1{#3 \DIFF[#2] #4}}

\NewDocumentCommand{\Update}{somm}{\Parens#1{#3 \UPDATE[#2] #4}}
\NewDocumentCommand{\Apply}{somm}{\Parens#1{#4 \UPDATE[#2] #3}}

\NewDocumentCommand{\WideDiff}{somm}{\Parens#1{#3 \;\;\DIFF[#2]\;\; #4}}
\NewDocumentCommand{\WideApply}{somm}{\Parens#1{#4 \;\;\APPLY[#2]\;\; #3}}
\NewDocumentCommand{\COMPOSE}{}{\circ}
\NewDocumentCommand{\Compose}{smm}{\Parens#1{#2 \COMPOSE #3}}

\NewDocumentCommand{\NIL}{}{\mathbf{0}}
\NewDocumentCommand{\Nil}{om}{\NIL_{#2}}

\NewDocumentCommand{\ChangeStruct}{m}{\widehat{#1}}

\NewDocumentCommand{\MAP}{}{\Term{map}}

\NewDocumentCommand{\mean}{soomo}{\Parens#1{%
\left\llbracket\,#4\,\right\rrbracket\IfNoValueTF{#5}{}{^{#5}}\IfNoValueTF{#2}{}{\;#2}\IfNoValueTF{#3}{}{\;#3}}}

\NewDocumentCommand{\update}{smmm}{\Parens#1{#4[#2\mapsto#3]}}

\NewDocumentCommand{\D}{m}{\Term{d#1}}

\NewDocumentCommand{\Old}{m}{#1_{\textrm{old}}}
\NewDocumentCommand{\New}{m}{#1_{\textrm{new}}}

\NewDocumentCommand{\APP}{}{\Keyword{app}}
\NewDocumentCommand{\IF}{}{\mathop{\Keyword{if}}}
\NewDocumentCommand{\THEN}{}{\mathbin{\Keyword{then}}}
\NewDocumentCommand{\ELSE}{}{\mathbin{\Keyword{else}}}
\NewDocumentCommand{\TRUE}{}{\Keyword{true}}
\NewDocumentCommand{\FALSE}{}{\Keyword{false}}
\NewDocumentCommand{\BAG}{}{\mathop{\Keyword{Bag}}}
\NewDocumentCommand{\UNION}{}{\Term{union}}
\NewDocumentCommand{\NAT}{}{\Keyword{Nat}}
\NewDocumentCommand{\INT}{}{\Keyword{Int}}
\NewDocumentCommand{\BOOL}{}{\Keyword{Bool}}
\NewDocumentCommand{\DERIVE}{}{\mathcal{Derive}}

\NewDocumentCommand{\NEGATE}{}{\Term{negate}}
\NewDocumentCommand{\CHANGE}{o}{\Delta\Index{#1}}
\NewDocumentCommand{\BASE}{}{\mathop{\Keyword{Base}}}

\NewDocumentCommand{\Parens}{mm}{\IfBooleanTF#1{\left(#2\right)}{#2}}

\NewDocumentCommand{\Fun}{smm}{\Parens#1{#2 \to #3}}
\NewDocumentCommand{\Base}{sO{}}{\Parens#1{\App\BASE{#2}}}
\NewDocumentCommand{\Bag}{sO{}}{\Parens#1{\IfValueTF{#1}{\App\BAG{#2}}{\BAG}}}
\NewDocumentCommand{\Nat}{s}{\Parens#1{\NAT}}
\NewDocumentCommand{\Int}{s}{\Parens#1{\INT}}

\NewDocumentCommand{\Set}{m}{\left\lbrace#1\right\rbrace}

\NewDocumentCommand{\App}{smm}{\Parens#1{#2\;#3}}
\NewDocumentCommand{\Lam}{smm}{\Parens#1{\lambda#2.\ #3}}
\NewDocumentCommand{\Var}{sm}{\Parens#1{#2}}
\NewDocumentCommand{\Const}{sm}{\Parens#1{#2}}
\NewDocumentCommand{\Lit}{sm}{\Parens#1{#2}}
\NewDocumentCommand{\Add}{smm}{\Parens#1{#2 + #3}}
\NewDocumentCommand{\Minus}{sm}{\Parens#1{- #2}}
\NewDocumentCommand{\Bool}{s}{\Parens#1{\BOOL}}
\NewDocumentCommand{\True}{s}{\Parens#1{\TRUE}}
\NewDocumentCommand{\False}{s}{\Parens#1{\FALSE}}
\NewDocumentCommand{\If}{smmm}{\Parens#1{\IF #2 \THEN #3 \ELSE #4}}
\NewDocumentCommand{\Empty}{s}{\Parens#1{\emptyset}}

\NewDocumentCommand{\SINGLETON}{}{\Term{singleton}}
\NewDocumentCommand{\Singleton}{sm}{\Parens#1{\App\SINGLETON{#2}}}
\NewDocumentCommand{\INSERT}{}{\Term{insert}}
\NewDocumentCommand{\Insert}{smm}{\Parens#1{\App{\App\INSERT{#2}}{#3}}}
\NewDocumentCommand{\Union}{smm}{\Parens#1{\App{\App\UNION {#2}}{#3}}}
\NewDocumentCommand{\Map}{smm}{\Parens#1{\App{\CMap{#2}}{#3}}}
\NewDocumentCommand{\CMap}{sm}{\Parens#1{\App{\MAP}{#2}}}
\NewDocumentCommand{\FLATMAP}{}{\Term{flatmap}}
\NewDocumentCommand{\Flatmap}{smm}{\Parens#1{\App{\App\FLATMAP{#2}}{#3}}}
\NewDocumentCommand{\Sum}{sm}{\Parens#1{\App{\Term{sum}}{#2}}}
\NewDocumentCommand{\Negate}{sm}{\Parens#1{\App{\NEGATE}{#2}}}
\NewDocumentCommand{\DELETE}{}{\Term{delete}}
\NewDocumentCommand{\Delete}{smm}
  {\Parens#1{\App{\App{\App{\DELETE}{#2}}{\Term{from}}}{#3}}}

\NewDocumentCommand{\Change}{som}{\Parens#1{\CHANGE[#2]#3}}

\NewDocumentCommand{\Eval}{m}{\mean{#1}}
\NewDocumentCommand{\EvalWith}{smm}{\Parens#1{\mean[#3]{#2}}}
\NewDocumentCommand{\EvalInc}{m}{\mean{#1}[\GD]}
\NewDocumentCommand{\EvalIncWith}{smmm}{\Parens#1{\mean[#3][#4]{#2}[\GD]}}

\NewDocumentCommand{\EvalIncSmashWith}{smmm}{\Parens#1{\mean[#3][#4]{#2}[\smash{\GD}\!\!\!]}}

\NewDocumentCommand{\EvalConst}{sm}{\Parens#1{\mean{#2}[\text{const}]}}
\NewDocumentCommand{\EvalIncConst}{sm}{\Parens#1{\mean{#2}[\text{$\GD$ const}]}}
\NewDocumentCommand{\EvalBase}{sm}{\Parens#1{\mean{#2}[\text{base}]}}
\NewDocumentCommand{\Derive}{sm}{\Parens#1{\DERIVE(#2)}}
\NewDocumentCommand{\DeriveConst}{sm}{\Parens#1{\DERIVE_{\text{const}}(#2)}}

\NewDocumentCommand{\Append}{smm}{\Parens#1{#2 , #3}}
\NewDocumentCommand{\HasType}{smm}{\Parens#1{#2 : #3}}

\NewDocumentCommand{\EmptyContext}{}{\varepsilon}
\NewDocumentCommand{\Extend}{sO{\Gamma}mm}{\Parens#1{\Append{#2}{\HasType{#3}{#4}}}}
\NewDocumentCommand{\Typing}{O{\Gamma}mm}{#1 \vdash \HasType{#2}{#3}}

\NewDocumentCommand{\HasValue}{smm}{\Parens#1{#2 = #3}}
\NewDocumentCommand{\EmptyEnv}{}{\varnothing}
\NewDocumentCommand{\ExtendEnv}{sO{\rho}mm}{\Parens#1{\Append{#2}{\HasValue{#3}{#4}}}}

\NewDocumentCommand{\Rule}{O{}mm}{\inferrule{#2}{#3}\;\textsc{#1}}
\NewDocumentCommand{\Axiom}{O{}m}{\Rule[#1]{\hbox{}}{#2}}

\NewDocumentCommand{\IMPLEMENTS}{oo}{\sim\Index{#1}\Index*{#2}}
\NewDocumentCommand{\Implements}{soomm}{\Parens#1{#4\IMPLEMENTS[#2][#3] #5}}

\NewDocumentCommand{\DeltaType}{m}{\GD #1}

\NewDocumentCommand{\DeltaContext}{m}{\GD #1}

\newenvironment{syntax}
{\[\begin{tabular}{>{$}r<{$}@{\;}>{$}c<{$}@{\;}>{$}l<{$}@{\qquad}l}}
{\end{tabular}\]}

\NewDocumentCommand{\FramedSignature}{m}
  {\fbox{\(#1\)}}

\NewDocumentCommand{\RightFramedSignature}{m}
  {\vbox{\hfill\FramedSignature{#1}}}

\NewDocumentEnvironment{typing}{o}
{\IfValueTF{#1}{\RightFramedSignature{#1}}{}
 \begin{mathpar}}
{\end{mathpar}}

\newenvironment{signature}
{\begin{tabular}{>{$}c<{$}@{$\hbox{}:\hbox{}$}>{$}l<{$}@{\qquad}l}}
{\end{tabular}}

\def\Case#1:{\smallbreak\noindent\textit{Case} $#1$:\par}

\newdimen\eqsep
\NewDocumentCommand{\HISTOGRAM}{}{\Term{histogram}}
\NewDocumentCommand{\WORDCOUNT}{}{\Term{wordcount}}
\NewDocumentCommand{\HASHMAP}{}{\Keyword{Map}}
\NewDocumentCommand{\HashMap}{mm}{\App{\App\HASHMAP{#1}}{#2}}
\NewDocumentCommand{\DOCUMENT}{}{\Keyword{Document}}
\NewDocumentCommand{\DOCUMENTID}{}{\Keyword{ID}}
\NewDocumentCommand{\WORD}{}{\Keyword{Word}}

\NewDocumentCommand{\FOLDBAG}{}{\Term{foldBag}}
\NewDocumentCommand{\FoldBag}{smm}{\Parens#1{\App{\App\FOLDBAG {#2}}{#3}}}

\NewDocumentCommand{\ABELIAN}{}{\Keyword{AbelianGroup}}
\NewDocumentCommand{\Abelian}{sm}{\Parens#1{\App\ABELIAN{#2}}}

\NewDocumentCommand{\FOLDMAP}{}{\Term{foldMap}}

\NewDocumentCommand{\MAYBE}{}{\Keyword{Maybe}}
\NewDocumentCommand{\Maybe}{sm}{\Parens#1{\App\MAYBE{#2}}}

\PrerenderUnicode{λ}
\DeclareRobustCommand{\TitleLambda}{\texorpdfstring{$\Gl$}{λ}}

\def\Gi{\iota   }

\def\Gl{\lambda }
\def\Gr{\rho    }
\def\Gs{\sigma  }
\def\Gt{\tau    }

\def\GG{\Gamma  }
\def\GD{\Delta  }

\usepackage{amsthm,listings,fixfoot}
\lstdefinelanguage{scala}{
  morekeywords={%
	  abstract,case,catch,class,def,do,else,extends,%
	  false,final,finally,for,forSome,if,implicit,import,lazy,%
	  match,new,null,object,override,package,private,protected,%
	  return,sealed,super,this,throw,trait,true,try,type,%
	  val,var,while,with,yield},
  otherkeywords={=>,<-,<\%,<:,>:,\#,@,scala>},
  sensitive=true,
  morecomment=[l]{//},
  morecomment=[n]{/*}{*/},
  morestring=[b]",
  morestring=[b]',
  morestring=[b]"""
}[keywords,comments,strings]

\usepackage[capitalise]{cleveref}

\crefname{section}{Sec.}{Sec.}
\crefname{subsection}{Sec.}{Sec.}
\crefname{subsubsection}{Sec.}{Sec.}

\theoremstyle{definition}
\usepackage{etoolbox}
\providecommand{\openbox}{\leavevmode
  \hbox to.77778em{%
  \hfil\vrule
  \vbox to.675em{\hrule width.6em\vfil\hrule}%
  \vrule\hfil}}
\makeatletter
\DeclareRobustCommand{\QED}{%
  \ifmmode
    \eqno \def\@badmath{$$}%
    \let\eqno\relax \let\leqno\relax \let\veqno\relax
    \hbox{\openbox}%
  \else
    \leavevmode\unskip\penalty9999 \hbox{}\nobreak\hfill
    \quad\hbox{\openbox}%
  \fi
}
\def\qedAligned{\tag*{\hbox{\openbox}}\gdef\qed{}}
\gdef\qed{\QED\gdef\qed{}}
\patchcmd{\@endtheorem}{\endtrivlist}{\qed\gdef\qed{\QED\gdef\qed{}}\endtrivlist}{}{}
\makeatother

\newtheorem{theorem}{Theorem}[section]
\newtheorem{lemma}[theorem]{Lemma}

\newtheorem{definition}[theorem]{Definition}

\newlist{subdefinition}{enumerate}{1}
\setlist*[subdefinition]{label=(\alph*), ref=\arabic{section}.\arabic{theorem}\alph*}
\crefname{subdefinitioni}{Property}{Properties}
\Crefname{subdefinitioni}{Property}{Properties}

\newlist{subparameter}{enumerate}{1}
\setlist*[subparameter]{label=(\alph*), ref=\arabic{section}.\arabic{theorem}\alph*}
\crefname{subparameteri}{Plugin Requirement}{Plugin Requirements}
\Crefname{subparameteri}{Plugin Requirement}{Plugin Requirements}

\newlist{subtheorem}{enumerate}{1}
\setlist*[subtheorem]{label=(\alph*), ref=\arabic{section}.\arabic{theorem}\alph*}
\crefname{subtheoremi}{Theorem}{Theorems}
\Crefname{subtheoremi}{Theorem}{Theorems}

\newcommand{\ILC}{I{\TitleLambda}C}

\begin{document}

\conferenceinfo{PLDI '14}{June 9--11, 2014, Edinburgh, UK}
\copyrightyear{2014}
\copyrightdata{[to be supplied]}

\title{A Theory of Changes for Higher-Order Languages}
\subtitle{Incrementalizing {\TitleLambda}-Calculi by Static Differentiation}

\iftoggle{names}{%
\newcommand{\unimarburg}{Philipps-Universität Marburg}
\authorinfo{Yufei Cai}
           {\unimarburg}{}
\authorinfo{Paolo G. Giarrusso}
           {\unimarburg}{}
\authorinfo{Tillmann Rendel}
           {\unimarburg}{}
\authorinfo{Klaus Ostermann}
           {\unimarburg}{}
}{%
\authorinfo{}
           {}{}
}

\maketitle

\begin{abstract}
If the result of an expensive computation is invalidated by a small
change to the input, the old result
should be updated incrementally instead of reexecuting the whole computation.
We incrementalize programs through
their \emph{derivative}. A derivative maps
changes in the program's input directly to changes in the program's
output, without reexecuting the original program. We present a
program transformation taking programs to their
derivatives, which is fully static and automatic, supports first-class
functions, and produces derivatives amenable to standard
optimization.

We prove the program transformation correct in Agda for a
family of simply-typed $\lambda$-calculi, parameterized by base
types and primitives. A precise interface specifies what is
required to incrementalize the chosen primitives.

We investigate performance by a case study: We implement in Scala the
program transformation, a plugin and improve performance of a nontrivial program by
orders of magnitude.

\end{abstract}

\section{Introduction}
\label{sec:intro}

Incremental computation has a long-standing history in computer
science~\citep{Ramalingam93}. Often, a program needs to update its
output efficiently to reflect input
changes~\citep{Salvaneschi13reactive}. Instead of rerunning such a
programs from scratch on its updated input, incremental
computation research looks for alternatives that are cheaper in a common scenario:
namely, when the input change is much smaller than the input itself.

\pg{Ensure this works as running example. In the end, we should
  also show its derivative (probably later).}
For instance, consider the following program which adds all members of a collection $s$ of numbers.
\begin{align*}
\Term{sum}\;s & = \Term{fold}\;(+)\;0\;s\\
\Var{y} & = \Term{sum}\;\Set{1, 2, 3, 4}
\end{align*}
Now assume that the input to $\Term{sum}$ changes from $\Set{1, 2, 3, 4}$ to $\Set{2, 3, 4, 5}$.
Instead of recomputing $y$ from scratch, we could also compute it incrementally. If we have a
representation for the change to the input (say, $\D s = \Set{\Keyword{remove} \; 1, \Keyword{add} \; 5}$), we can compute the new
result through a function $\Term{sum}'$ that takes the old input $s = \Set{1, 2, 3, 4}$ and the change $\D s$ and produces a change $\D y$ to the
output $y$. In this case, it would compute the change $\D y = \Term{sum}' \; s \; \D s = \Keyword{plus} \; 4$, which can then be used to update the original output $y = 10$
to yield the updated result $14$. We call $\Term{sum}'$ the \emph{derivative} of $\Term{sum}$.
It is a function in the
same language of $\Term{sum}$, accepting and producing changes, which
are simple first-class values of this language.
If we increase the size of the original input $s$, the
complexity of $\Term{sum}\;s$ increases linearly, while the complexity
of $\Term{sum}' \; s \; \D s$ only depends on the size of $\D s$,
which is smaller both in our example and typically.

To address this problem, in this paper we introduce the \ILC\
(incrementalizing $\Gl$-calculi) framework. We define
an automatic program transformation $\DERIVE$
that \emph{differentiates} programs, that is, computes their
derivatives; $\DERIVE$ guarantees that
\begin{equation}
  \label{eq:correctness}
\App{f}{\Apply*{\D a}{a}}
\cong
\Apply{\App*{\App{\Derive{f}}{a}}{\D a}}{\App*{f}{a}}.
\end{equation}
where
$\cong$ is denotational equality,
$\D a$ is a change on $a$ and $\Apply{\D a}{a}$ denotes $a$
updated with change $\D a$, that is, the updated input of $f$.
Hence, we can optimize programs by replacing the left-hand side,
which recomputes the output from scratch, with the right-hand
side, which computes the output incrementally using derivatives.

\ILC\ is based on a simply-typed $\Gl$-calculus
parameterized by \emph{plugins}. A plugin
defines
(a) base types and primitive operations, and
(b) a change representation for each base type, and an
incremental version for each primitive. In other words, the plugin
specifies the primitives and their respective derivatives, and
\ILC\ can glue together these simple derivatives in such a way
that derivatives for arbitrary simply-typed $\Gl$-calculus expressions
using these primitives can be computed. Both our implementation and our correctness proof 
is parametric in the plugins, hence it is easy to support (and prove correct)
new plugins.

This paper makes the following contributions:
\begin{itemize}
\item We present a novel mathematical theory of changes and derivatives, which is more
  general than other work in the field because changes are
  first-class entities, they are distinct from base values and
  they are defined also for functions (\cref{sec:1st-order-changes}).
\item We present the first approach to incremental computation for
pure $\lambda$-calculi by a source-to-source transformation, $\DERIVE$, that requires no run-time
support. The transformation produces an incremental program in the same language;
all optimization techniques for the original program are
applicable to the incremental program as well.
We prove that our incrementalizing transformation $\DERIVE$
is correct~(\cref{eq:correctness})
by a machine-checked formalization in Agda~\citep{agda-head}.
The proof gives insight into the definition of $\DERIVE$: we
first construct the derivative $\EvalInc{-}$ of the denotational
semantics of a simply-typed $\lambda$-calculus term, that is, its
\emph{change semantics}.
Then, we show that $\DERIVE$ is produced by erasing
$\EvalInc{-}$ to a simply-typed program (\cref{sec:correctness}).

\item While we focus mainly on the theory of changes
and derivatives, we also provide an initial experimental evaluation.
We implement the derivation transformation in Scala. The implementation
is organized as a plug-in architecture that can be extended with new base
types and primitives. We define a plugin with support for
different collection types and use the plugin to 
incrementalize a variant of the MapReduce programming model~\citep{Lammel07}.
  Benchmarks show that
  incrementalization can reduce asymptotic complexity and can turn $O(n)$
  performance into $O(1)$, improving running time by over 4
  orders of magnitude (\cref{sec:applying}).

\end{itemize}

Our Agda formalization, Scala implementation and benchmark
results are available at the URL
\url{https://www.dropbox.com/sh/3vg8pikd6wbgck5/SaZPRvqB2p}.
All lemmas and theorems presented
in this paper have been proven in Agda.
In the paper, we present an overview of
the formalization in more human-readable form, glossing over some
technical details.

\pg{Citations lost from the old intro:
  There was also \citep{GiarrussoAOSD13}, but for lack of space
  we should omit it for now.}

\section{A theory of changes}
\label{sec:1st-order-changes}

This section introduces a formal concept of changes; this
concept was already used informally in \cref{eq:correctness} and is central
to our approach. We first define change structures formally, then construct 
change structures for functions between change structures,
and conclude with a theorem that relates function changes to derivatives. 

\subsection{Change structures}\label{ssec:change-structures}
Consider a set of values, for instance the set of natural numbers
$\mathbb{N}$. A change $\D v$ for $v \in \mathbb{N}$ should
describe the difference between $v$ and another natural $\New{v}
\in \mathbb{N}$. We do not define changes directly, but we
specify operations which must be defined on them. They are:
\begin{itemize}
\item We can \emph{update} a base value $v$ with a
  change $\D v$ to obtain an updated or \emph{new} value
  $\New{v}$. We write $\New{v} = \Apply{\D v}{v}$.
\item We can compute a change between two arbitrary
  values $\Old{v}$ and $\New{v}$ of the set we are considering.
  We write $\D v = \Diff{\New{v}}{\Old{v}}$.
\end{itemize}

For naturals, it is usual to describe changes using standard
subtraction and addition. That is, for naturals we can define
$\Apply{\D v}{v} = v + \D v$ and $\Diff{\New{v}}{\Old{v}} =
\New{v} - \Old{v}$. To ensure that $\APPLY$ and $\DIFF$ are
always defined, we need to define the set of changes carefully.
$\mathbb{N}$ is too small, because subtraction does not always
produce a natural; the set of integers $\mathbb{Z}$ is instead
too big, since adding a natural and an integer does not always
produce a natural. In fact, we cannot use the same set of all
changes for all naturals. Hence we must adjust the requirements:
for each base value $v$ we introduce a set $\Change{v}$ of
changes for $v$, and require $\Diff{\New{v}}{\Old{v}}$ to produce
values in $\Change{\Old{v}}$, and $\Apply{\D v}{v}$ to be defined
for $\D v$ in $\Change{v}$. For natural $v$, we set $\Change{v} =
\left\{\D v \mid v + \D v \geq 0 \right\}$; $\DIFF$ and $\APPLY$ are
then always defined.

\pg{We never say why we use ``structure''. On second thought,
  this might be OK since we have little space.}
The following definition sums up the discussion so far:

\pg{Consider less heavyweight phrasing, such as: ``To each $v \in V$
  we associate a set of changes $\Change{v}$. But do this consistently.}
\begin{definition}[Change structures]
  \label{def:change-struct}
  A quadruple $\ChangeStruct{V} = (V, \CHANGE,
  \UPDATE,
  \DIFF)$ is a \emph{change structure} (for $V$) if the
  following holds.

  \begin{subdefinition}
  \item $V$ is a set.
  \item Given $v \in V$, $\Change{v}$ is a set, called the \emph{change set}.
  \item Given $v \in V$ and $\D v \in \Change{v}$, $\Apply{\D v}{v} \in V$.
    \label{def:update}
  \item Given $u, v \in V$, $\Diff{u}{v} \in \Change{v}$.
    \label{def:diff}
  \item Given $u, v \in V$, $\Apply{\Diff*{u}{v}}{v}$ equals $u$.
    \qed
    \label{def:update-diff}
  \end{subdefinition}
\end{definition}

We overload operators $\CHANGE$, $\DIFF$ and $\APPLY$ to refer
to the corresponding operations of different change structures;
we will subscript these symbols when needed to prevent ambiguity.
For any $\ChangeStruct{S}$, we write $S$ for its first component,
as above.

One might expect a further assumption that
$\Diff{\Apply*{\D v}{v}}{v} = \D v$. While it does hold
for the change structure of $\mathbb{N}$, it is not needed in general.
This means that multiple changes can represent the difference between
the same two base values. Throughout our theory, we only discuss equality of
base values, not of changes.

\paragraph{Examples.}
One way to define change structures is from abelian groups. In
algebra, an abelian group is a quadruple $(G, \boxplus,
\boxminus, e)$, where $\boxplus$ is a commutative
and associative binary operation, $e$ is its identity
element, and $\boxminus$ produces inverses of elements $g$
of $G$, such that $(\boxminus g) \boxplus g = g \boxplus
(\boxminus g) = e$. For instance, integers,
unlike naturals, form the abelian group $(\mathbb{Z}, +, -, 0)$
(where $-$ represents the unary minus). Each abelian group
$(G, \boxplus, \boxminus, e)$ induces a change structure,
namely $\left(G, \Lam{g}{G}, \boxplus, \Lam{g\; h}{g
    \boxplus (\boxminus h)}\right)$, where the change set
for any $g \in G$ is the whole $G$. Change structures
are more general, though, as the example with natural numbers illustrates.

The abelian group on integers induces also a change structure on
integers, namely $\ChangeStruct{\mathbb{Z}} = (\mathbb{Z},
\Lam*{v} {\mathbb{Z}}, +, -)$, where $\DIFF$ and $\UPDATE$ have
the same definitions as for naturals.

Another useful example is the definition of an abelian group (and
the induced change structure) on
bags with signed multiplicities~\citep{Koch10IQE}. These are
unordered collections where each element can appear an integer
number of times. Element can appear a negative number $-n$ of
times in a bag change to represent $n$ removals of that element.
If $\Empty$ represents the empty bag,
$\UNION$ performs bag union, and $\NEGATE$ negates the multiplicities of elements, we can define
the abelian group $(\Bag[\Gi], \UNION, \NEGATE, \Empty)$,
which induces the change structure
$\ChangeStruct{\Bag[\Gi]} = (\Bag[\Gi], \Lam*{v}
{\Bag[\Gi]}, \UNION,$ $\Lam{x\; y}{\Union{x}{\Negate*{y}}})$.

\paragraph{Nil changes and derivatives.}
A particularly important change is the \emph{nil change} of a value:
\begin{definition}[Nil change]
  \label{def:nil-change}
  Given a change structure $\ChangeStruct{V}$ and a value $v \in V$, the change
  $\Diff{v}{v}$ is the nil change for $v$.
  \[
    \Nil{v} = \Diff{v}{v} \qed
  \]
\end{definition}
The nil change for a value does indeed not change it.
\begin{lemma}[Behavior of $\NIL$]
  \label{thm:update-nil}
  Given a change structure $\ChangeStruct{V}$ and a value $v \in V$,
  $\Apply{\Nil{v}}{v} = v$.
\end{lemma}

\begin{optionalproof}
Follows from \cref{def:update-diff,def:nil-change}.
\end{optionalproof}

Equipped with the preceding definition, we can now restate the definition of derivatives from \cref{eq:correctness}.

\begin{definition}[Derivatives]
  \label{def:derivatives}
  Given change structures $\ChangeStruct{A}$ and $\ChangeStruct{B}$ and a function $f \in A \to
  B$ on the change sets of these change structures, we call a binary function $f'$ the \emph{derivative} of $f$ if
  for all values $a \in A$ and corresponding changes $\D a \in
  \Change[A]{a}$,
  \[\App{f}{\Apply*[A]{\D a}{a}} = \Apply[B]{\App*{\App{f'}{a}}{\D a}}{\App*{f}{a}}\text{.}\qed\]
\end{definition}

To avoid parentheses, we give function application precedence over
$\oplus$ and $\ominus$ in the remainder of this paper. For instance, the equation above can be
written as 
$\App{f}{\Apply*[A]{\D a}{a}} = \Apply[B]{\App{\App{f'}{a}}{\D a}}{\App{f}{a}}$.

\subsection{Function changes}
\label{sec:function-change}

\begin{figure*}
\subfigure[Syntax.\label{fig:syntax}]{
\parbox[b]{0.40\linewidth}{
\begin{tabular}{>{$}r<{$}@{$\;::=\;$}>{$}c<{$}@{$\;$}>{$}l<{$}@{\quad}>{(}l<{)}}
\Gi      & \rlap{\ldots} &                       & base types\\
\Gs, \Gt & \Gi           & \mid \Fun{\Gt}{\Gt}   & types\\
\GG      & \EmptyContext & \mid \Extend{x}{\tau} & typing contexts\\
c        & \rlap{\ldots} &                       & constants\\
s, t     & c             & \mid \Lam{x}{t}
                           \mid \App{t}{t}
                           \mid x                & terms
\end{tabular}
\vspace{0.5\baselineskip}
}}\hfill
\subfigure[Typing.\label{fig:typing}]{
\parbox[b]{0.58\linewidth}{
\begin{typing}
\noindent
\Rule[Const]
  {\ldots}
  {\Typing[]{c}{\tau}}

\Axiom[Lookup]
  {\Typing[\Append{\GG_1}{\Append{\HasType{x}{\tau}}{\GG_2}}]{\Var{x}}{\tau}}

\raisebox{0.5\baselineskip}{\fbox{$\Typing{t}{\tau}$}}

\Rule[Lam]
  {\Typing[\Extend{x}{\Gs}]{t}{\Gt}}
  {\Typing{\Lam{x}{t}}{\Fun{\Gs}{\Gt}}}

\Rule[App]
  {\Typing{s}{\Fun{\Gs}{\Gt}}\\
   \Typing{t}{\Gs}}
  {\Typing{\App{s}{t}}{\Gt}}
\end{typing}
\vspace{-0.25\baselineskip}
}}
\caption{Our base calculus.}
\label{fig:base-calculus}
\end{figure*}

We will now demonstrate that we can construct change structures
for functions between change structures.

A higher-order function $f$ can take other functions as
arguments or return them as results. Hence, the derivative of $f$
will respectively take function changes as arguments or return
function changes as results.
For instance, $f = \Lam x {\Lam y {x + y}}$ is a higher-order
function, so its derivative gives us the change to the function
$g = \Lam y {x + y}$ in terms of $x$ and its change $\D x$. 

\pg{Tillmann's plan, last time we talked, was to use change
  structures in the next equation, to avoid the abuse of
  notation.}

The first important design decision is how to represent changes to functions.
If a function has type $\Fun* \Gs \Gt$, we represent a change to that function
by a function of type $\Fun{\Gs}{\Fun{\Change\Gs}{\Change\Gt}}$. By syntactically
abusing $\Delta$ as a type operator, we can write this as:
\begin{equation}
\label{eq:conflation-intro}
\Change{\Fun* \Gs \Gt} = \Fun{\Gs}{\Fun{\Change\Gs}{\Change\Gt}}.
\end{equation}

A function change $\D f$ hence takes as input the original value
$a$ and its change $\D a$. Once we define change structures for
functions, we will show that a function change produces as output
the difference between the updated output $\App {\Update*{f}{\D f}}
{\Update*{a}{\D a}}$ and the original output $\App f a$. This
difference is caused by two changes: the change to $a$ given by
$\D a$ and the change of $f$ itself given by $\D f$. \pg{Maybe add
  one sentence to highlight the importance of this conflation?}

We now define the set of function changes for function $f \in A
\to B$. To fulfill the definition of change structure
(\cref{def:change-struct}), function changes must produce valid
changes for their codomain; moreover, it must be possible to ``flip''
an element change $\D a$ from a function change to its associated
function:

\begin{definition}
  \label{def:function-changes:change}
  Given change structures $\ChangeStruct{A}$ and $\ChangeStruct{B}$,
  the set $\Change[A \to B]{f}$ contains all binary functions $\D
  f$ so that
  \NewDocumentCommand{\TheNewValue}{}{\Apply*[A]{\D a}{a}}
  \begin{subdefinition}
    \item
      \label{def:function-changes:signature}
      $\App{\App{\D f}{a}}{\D a} \in \Change[B]{\App*{f}{a}}$ and
    \item
      \label{def:function-changes:validity}
      $\Apply[B]{\App{\App{\D f}{a}}{\D a}}{\App{f}{a}} =
      {\App{f}{\TheNewValue}}
      \UPDATE_B
      \App{\App{\D f}{\TheNewValue}}{\Nil{\TheNewValue}}$
  \end{subdefinition}
  for all values $a \in A$ and corresponding changes $\D a \in
  \Change[A]{a}$.
\end{definition}

The change structure operations on functions can now be defined as a distributive law.

\begin{definition}[Operations on function changes]
  \label{def:function-changes:update}
  \label{def:function-changes:diff}
  Given change structures $\ChangeStruct{A}$ and $\ChangeStruct{B}$,
  the operations $\APPLY[A \to B]$ and $\DIFF[A \to B]$ are
  defined as follows.
  \begin{alignat*}{5}
    &\App{(\Update[A \to B]{f&&}{\D f})}{&&v}
      && = \Update[B]{\App{f}{v}&&}{\App{\App{\D f}{v}}{\Nil[A]{v}}}\\
    &\App{\App{(\Diff[A \to B]{f_2&&}{f_1})}{&&v}}{\D v}
      && = \Diff[B]{\App{f_2}{\Update*[A]{v}{\D v}}&&}{\App{f_1}{v}}\qedAligned
  \end{alignat*}
\end{definition}

\begin{optionallemma}
  \label{thm:diff-valid}
  Given change structures $\ChangeStruct{A}, \ChangeStruct{B}$ and functions $f_1, f_2 \in A
  \to B$, then $\Diff[A \to B]{f_2}{f_1} \in \Change[A \to B]{f_1}$.
\end{optionallemma}

\begin{optionalproof}
  We have to verify the two properties of
  \cref{def:function-changes:change}. The first follows from
  \cref{def:diff} for the change structure $\ChangeStruct{B}$. It remains to
  verify \cref{def:function-changes:validity}.

  Let $a_1 \in A$ be an arbitrary value with a corresponding
  change $\D a \in \Change[A]{a}$, and let $a_2$ be
  $\Apply{\D a}{a_1}$, then
  \begin{align*}
  & \Apply[B]
      {\App{\App{\Diff*[A \to B]{f_2}{f_1}}{a_1}}{\D a}}
      {\App{f_1}{a_1}}\\
  & \quad = \Apply[B]
               {\Diff*[B]
                 {\App{f_2}{a_2}}
                 {\App{f_1}{a_1}}}
               {\App{f_1}{a_1}}\\
  & \quad = \App{f_2}{a_2}\\
  & \quad = \Apply[B]
              {\Diff*[B]
                {\App{f_2}{a_2}}
                {\App{f_1}{a_2}}}
              {\App{f_1}{a_2}}\\
  & \quad = \Apply[B]
              {\Diff*[B]
                {\App{f_2}{\Apply*{\Nil[B]{a_2}}{a_2}}}
                {\App{f_1}{a_2}}}
              {\App{f_1}{a_2}}\\
  & \quad = \Apply[B]
              {\App{\App{\Diff*[A \to B]{f_2}{f_1}}{a_2}}{\Nil{a_2}}}
              {\App{f_1}{a_2}}
  \end{align*}
  by
  \cref{def:function-changes:diff,def:update-diff,thm:update-nil}.
\end{optionalproof}

All these definitions have been carefully set up to ensure that we have
in fact lifted change structures to function spaces.

\begin{theorem}
  \label{thm:func-changestruct}
  Given change structures $\ChangeStruct{A}$ and $\ChangeStruct{B}$, the quadruple $(A \to B, \CHANGE[A
  \to B], \UPDATE[A \to B], \DIFF[A \to B])$ is a
  change structure, which we denote by $\ChangeStruct{A} \to \ChangeStruct{B}$.
\end{theorem}

\begin{optionalproof}
  We have to verify the five properties of
  \cref{def:change-struct}. The first two follow by
  construction. \Cref{def:update} follows from the corresponding
  property of the change structure $\ChangeStruct{B}$. \Cref{def:diff} is
  verified in \cref{thm:diff-valid}. It remains to verify
  \cref{def:update-diff}.

  Let $f_1, f_2 \in A \to B$ be arbitrary functions. We show that
  $\Apply[A \to B]{\Diff*[A \to B]{f_2}{f_1}}{f_1}$ is
  extensionally equal to $f_2$. Let $a \in A$ be an arbitrary
  value, then
  \begin{align*}
    & \App{\Apply*[A \to B]{\Diff*[A \to B]{f_2}{f_1}}{f_1}}{a}\\
    & \quad = \Apply[B]
                {\App{\App{\Diff*[A \to B]{f_2}{f_1}}{a}}{\Nil[A]{a}}}
                {\App{f_1}{a}}\\
    & \quad = \Apply[B]
                {\Diff*[B]{\App{f_2}{\Apply*[A]{a}{\Nil[A]{a}}}}{\App{f_1}{a}}}
                {\App{f_1}{a}}\\
    & \quad = \Apply[B]
                {\Diff*[B]{\App{f_2}{a}}{\App{f_1}{a}}}
                {\App{f_1}{a}}\\
    & \quad = \App{f_2}{a}
  \end{align*}
  by the definitions of $\APPLY[A \to B]$ and $\DIFF[A \to B]$,
  \cref{thm:update-nil} for the change structure $\ChangeStruct{A}$ and
  \cref{def:update-diff} for the change structure $\ChangeStruct{B}$.
\end{optionalproof}

As promised, we can show that a function change $\D f$ reacts to
input changes $\D a$ like the incremental version of $f$, that is,
$\App{\App{\D f}{a}}{\D a}$ computes the change from
$\App{f}{a}$ to
$\App{\Apply*{\D f}{f}}{\Apply*{\D a}{a}}$:

\begin{lemma}[Incrementalization]
  \label{thm:incrementalization}
  Given change structures $\ChangeStruct{A}$ and $\ChangeStruct{B}$, a function $f \in A \to B$
  and a value $a \in A$ with corresponding changes $\D f \in
  \Change[A \to B]{f}$ and $\D a \in \Change[A]{a}$, we have that
  \[\App{\Apply*[A \to B]{\D f}{f}}{\Apply*[A]{\D a}{a}}
  = \Apply[B]{\App{\App{\D f}{a}}{\D a}}{\App{f}{a}}\text{.}\qed\]
\end{lemma}
The lemma is just a restatement of \cref{def:function-changes:validity}, which
  uses $\UPDATE$ on functions as defined in \cref{def:function-changes:update}.

\begin{optionalproof}
  \NewDocumentCommand{\TheNewValue}{}{\Apply*[A]{\D a}{a}}

  Let $f$, $a$, $\D f$ and $\D a$ be arbitrary, as in the statement. Then
  \begin{align*}
    & \App{\Apply*[A \to B]{\D f}{f}}{\Apply*[A]{\D a}{a}}\\
    & \quad = \Apply[B]{\App{\App{\D f}{\TheNewValue}}{\Nil{\TheNewValue}}}{\App{f}{\TheNewValue}}\\
    & \quad = \Apply[B]{\App{\App{\D f}{a}}{\D a}}{\App{f}{a}}
  \end{align*}
  by
  \cref{def:function-changes:update,def:function-changes:validity}
  as required.
\end{optionalproof}

For instance,
incrementalizing
\[
\APP = \Lam{f}{\Lam{x}{\App f x}}
\]
with respect to the input changes $\D f$, $\D x$ amounts to
calling $\D f$ on the original second argument $\Old x$ and on
the change $\D x$.

\subsection{Nil changes are derivatives}

\cref{thm:incrementalization} tells us about the form an
incremental program may take. If $\D f$ doesn't change $f$
at all, that is, if
$
\Apply{\D f}{f}= f
$,
then \cref{thm:incrementalization} becomes
\[
 \App {f} {\Apply* {\D a} {a}}
 =
\Apply {\App {\App {\D f} {a}} {\D a}} {\App{f}{a}}.
\]
It says that $\D f$ computes the change upon the output of $f$ 
given a change $\D a$ upon the input $a$ of $f$. In
other words, the nil change to a function is exactly its
derivative (see \cref{def:derivatives}):

\begin{theorem}[Nil changes are derivatives]
  \label{thm:nil-is-derivative}
  Given change structures $\ChangeStruct{A}$ and $\ChangeStruct{B}$ and a function $f \in A \to B$,
  the change $\Nil[A \to B]{f}$ is the derivative $f'$ of $f$.
\end{theorem}

\begin{optionalproof}
  Let $a \in A$ be an arbitrary value with a corresponding change
  $\D a \in \Change[A]{a}$. Then
  \begin{align*}
    & \App{f}{\Apply*[A]{\D a}{a}}\\
    & \quad = \App{\Apply*[A \to B]{\Nil[A \to B]{f}}{f}}{\Apply*[A]{\D a}{a}}\\
    & \quad = \Apply[B]{\App{\App{\Nil[A]{f}}{a}}{\D a}}{\App{f}{a}}
  \end{align*}
  holds by \cref{thm:update-nil,thm:incrementalization}, as
  required for derivatives by \cref{def:derivatives}.
\end{optionalproof}

In this section, we developed the theory of changes to define
formally what a derivative is (\cref{def:derivatives}) and to
recognize that in order to find the derivative of a function, we
only have to find its nil change
(\cref{thm:nil-is-derivative}). Next, we want to provide a fully
automatic method for finding the nil change of a given function.

\section{Incrementalizing \TitleLambda{}-calculi}
\label{sec:differentiate}
\label{sec:correctness}

In this section, we show how to incrementalize an arbitrary
program in simply-typed $\Gl$-calculus. To this end, we define
the source-to-source transformation $\DERIVE$. Using the
denotational semantics $\Eval{-}$ we define later (in
\cref{sec:denotational-sem}), we can specify $\DERIVE$'s intended
behavior: to ensure \cref{eq:correctness},
$\Eval{\Derive{f}}$ must be the derivative of $\Eval{f}$
for any closed term $\HasType{f}{A \to B}$. We will overload the word
``derivative'' and say simply that $\Derive{f}$ is the derivative of
$f$.

It is easy to define derivatives of arbitrary functions as:
\[\App{\App{f'}{x}}{\D x} = \Diff{\App{f}{\Update*{x}{\D x}}}{\App f x}\text{.}\]
We could implement $\DERIVE$ following the same strategy.
However, the resulting incremental programs would be no faster
than recomputation. We cannot do better for arbitrary mathematical functions,
since they are infinite objects which we cannot fully inspect.
Therefore, we resort to a source-to-source transformation
on simply-typed $\Gl$-calculus as defined in 
\cref{fig:base-calculus}. The sets of base types and primitive
constants, as well as the typing rules for primitive constants, are
on purpose left unspecified and only defined by plugins --- they are \emph{extensions points}.
Defining different plugins allows to experiment with
sets of base types, associated primitives and incrementalization strategies.
We show an example plugin in our case study
(\cref{sec:plugins}). In this section, we focus on the
incrementalization of the features that are shared among all
instances of the plugin interface, that is, function types and the
associated syntactic forms, $\Gl$-abstraction, application and
variable references.
\pg{Change this sentence if we move the plugin requirements.}
Throughout the section, we collect
requirements on the plugins that instantiate the
framework.
Definitions provided by the plugin are replaced, in figures, by ellipses
(``$\ldots$'').
Satisfying these requirements is sufficient to ensure
correct incrementalization.
\pg{Should we anticipate here that plugin requirements are
  collected at the end?}

\subsection{Change types and erased change structures}
\label{ssec:change-types}

\begin{figure}
\begin{signature}
\CHANGE  & \Fun{\ast}{\ast}
         & the type of changes\\[0.5ex]
\APPLY   & \Fun{\tau}{\Fun{\Change{\tau}}{\tau}}
         & update a value with a change\\
\DIFF    & \Fun{\tau}{\Fun{\tau}{\Change{\tau}}}
         & the change between two values\\
\end{signature}
\caption{Erased change structures on simple types.}
\label{fig:change-operations}
\end{figure}

\begin{figure}
\begin{align*}
\Change{\Fun* \Gs \Gt} &= \Fun{\Gs}{\Fun{\Change\Gs}{\Change\Gt}}\\
\DIFF_{\Fun{\Gs}{\Gt}} & = \Lam{g\ f\ x\ \D x}
  {\Diff{\App*g{\Apply*{\D x}{x}}}{\App*f x}}\\
\APPLY_{\Fun{\Gs}{\Gt}} & = \Lam{f\ \D f\ x}
  {\Apply{\App*{\App{\D f}{x}}{\Diff*xx}}{\App*f x}}
\end{align*}
\caption{The erased change structures for function types.}
\label{fig:diff-apply}
\end{figure}

We developed the theory of change structures in the previous
section to guide our implementation of $\DERIVE$. By
\cref{thm:nil-is-derivative}, $\DERIVE$ has only to find the nil
change to the program itself, because nil changes \emph{are}
derivatives. However, the theory of change structures is not
directly applicable to the simply-typed $\Gl$-calculus, 
because a precise implementation of
change structures requires dependent types. For instance,  
we cannot describe the set of
changes $\Change[\Gt]{v}$ precisely as a non-dependent type, because it depends on the value we plan
to update with these changes.

To work around this limitation of
our object language, we use a form of \emph{erasure} of dependent types
to simple types. In \cref{fig:change-operations} and \cref{fig:correctness:change-types}, we
define change types $\Change{\Gt}$ as an approximate description
of change sets $\Change[\Gt]{v}$ (\cref{fig:correctness:changes}). 
In particular, all changes in $\Change[\Gt]{v}$ correspond to values of terms with type $\Change{\Gt}$,
but not necessarily the other way around. 
For instance, in the
change structure for natural numbers described in \cref{ssec:change-structures}, we would
have $\Change{\Nat} = \Int$, even though not every
integer is a change for every natural number.
For primitive types $\iota$, 
$\Change{\iota}$ and its associated $\oplus$ and $\ominus$ operator
must be provided by the plugin developer.
For function types, erased change structures are given by \cref{fig:diff-apply}.
Erasing dependent types in all components of a change structure,
we obtain \emph{erased change structures}, which represent change
structures as simply-typed $\Gl$-terms
where $\UPDATE$ and $\DIFF$ are
families of $\Gl$-terms. 

Erased change structures are not change structures themselves.
However, we will show how change structures and erased changes
structures have ``almost the same'' behavior
(\cref{sec:differentiate-correct}). We will hence be able to
apply our theory of changes.

\subsection{Differentiation}
\label{ssec:differentiation}

When $f$ is a closed term of function type,
$\Derive{f}$ should be its derivative. More in general, as discussed, we want
that when $t$ is a closed term, $\Derive{t}$ is its nil change.
Since $\DERIVE$ recurses on open terms, we need a more general
specification.
We require that if $\Typing{t}{\tau}$, then $\Derive{t}$
represents the change in $t$ (of type $\Change{\Gt}$) in terms of
changes to the values of its free variables. As a special case,
when $t$ is a closed term, there is no free variable to change;
hence, the change to $t$ will be as desired the nil change of
$t$.

The following typing rule shows the static semantics of
$\DERIVE$:
\begin{typing}
\Rule[Derive]
  {\Typing{t}{\tau}}
  {\Typing[\Append{\GG}{\DeltaContext{\GG}}]{\Derive{t}}{\DeltaType{\tau}}}
\end{typing}

We see that $\Derive{t}$ has access both to the
free variables in $t$ (from $\GG$) and to their changes (from
$\DeltaContext{\GG}$, defined in
\cref{fig:correctness:change-contexts}).
For example, if a
well-typed term $t$ contains $x$ free, then $\GG$ contains an
assumption $\HasType{x}{\Gt}$ for some $\Gt$ and
$\DeltaContext{\GG}$ contains the corresponding assumption
$\HasType{\D x}{\DeltaType{\Gt}}$. Hence, $\Derive{t}$ can
access the change of $x$ by using $\D x$. For simplicity, 
we assume that the original program contains no variable names
that start with $\D{}$.%
The definition of $\DERIVE$ will ensure that
the $\D x$ variables are bound if the original term is closed.

Let us analyzes each case of the definition of $\Derive{u}$
(\cref{fig:correctness:derive}):
\begin{itemize}
\item If $u = x$, $\Derive{x}$ must be the change of $x$, that is
$\D x$.
\item If $u = \Lam{x}{t}$, $\Derive{t}$ is the change of
  $u$ given the change in its free variables. The change of $u$
  is then the change of $t$ as a function of the \emph{base input}
  $x$ and its change
  $\D x$, with respect to changes in other open variables. Hence,
  we simply need to bind $\D x$ by defining $\Derive{\Lam{x}{t}}
  = \Lam{x}{\Lam{\D x}{\Derive{t}}}$.
\item If $u = \App{s}{t}$, $\Derive{s}$ is the change of $s$ as a function
  of its base input and change. Hence, we simply apply $\Derive{s}$ to 
  the actual base input $t$ and change $\Derive{t}$, giving
  $\Derive{\App{s}{t}} =
  \App{\App{\Derive{s}}{t}}{\Derive{t}}$.
\item If $t = c$: since $c$ is a closed term, its
  change is a nil change, hence (by \cref{thm:nil-is-derivative}) $c$'s derivative. We can
  obtain a correct derivative for constants by setting:
  \[
  \Derive{c} =
  \Diff{c}{c} = \Nil{c} = c'
  \]
  This definition is inefficient for functional constants; hence plugins must provide derivatives
  of the primitives they define.
\end{itemize}

This might seem deceptively simple. But $\lambda$-calculus only
implements binding of values, leaving ``real work'' to
primitives; likewise, differentiation for $\lambda$-calculus only
implement binding of changes, leaving ``real work'' to
derivatives of primitives.
However, our support for
$\Gl$-calculus allows to \emph{glue} the primitives together.

We have now informally
derived the definition of $\DERIVE$ (\cref{fig:correctness:derive})
from its specification (\cref{eq:correctness}) and
its typing. But formally speaking,
we have defined $\DERIVE$, hence we must 
prove that $\DERIVE$ satisfies \cref{eq:correctness}. This proof
is discussed in the remainder of the section.

\begin{figure*}
  \small

  \NewDocumentCommand{\Subcaption}{mm}
    {\subfigure[\label{#1}{#2}]{\rule{\linewidth}{0pt}}\vspace{0.8cm}}

  \NewDocumentCommand{\Align}{m}
    {{\begin{align*}#1\end{align*}}\vspace*{-0.8cm}}

  \centering
  \begin{tabular}{p{0.25\linewidth}p{0.40\linewidth}p{0.25\linewidth}}
    \hfill
    \FramedSignature{\Change{\Gt}}
    &
    \hfill
    \FramedSignature{\D v, \D f \in \Change[\Gt]{v}}
    &
    \hfill
    \FramedSignature{v, f \in \Eval{\Gt}}
    \\
    \Align{
      \Change{\iota}
        & = \ldots\\
      \Change{\Fun*{\Gs}{\Gt}}
        & = \Gs \to
            \Change{\Gs} \to
            \Change{\Gt}
    }
    &
    \Align{
      \Change[\iota]{v}
        & = \ldots \subseteq \Eval{\Change{\Gi}}\\
      \Change[\Fun*{\Gs}{\Gt}]{f}
        & = \left\{ \D f \in \HasType*{x}{\Eval{\Gs}} \to
            \Change[\Gs]{x} \to
            \Change[\Gt]{\App*{f}{x}} \mid \right.\\
        & \left.\qquad
          \App{\Apply*[A \to B]{\D f}{f}}{\Apply*[A]{\D a}{a}}
          = \Apply[B]{\App{\App{\D f}{a}}{\D a}}{\App{f}{a}}
        \right\}
    }
    &
    \Align{
      \Eval{\iota}
        & = \ldots\\
      \Eval{\Fun{\Gs}{\Gt}}
        & = \Eval{\Gs} \to \Eval{\Gt}
    }
    \\
    \Subcaption{fig:correctness:change-types}{
      Change types.
    }
    &
    \Subcaption{fig:correctness:changes}{
      Change values.
    }
    &
    \Subcaption{fig:correctness:values}{
      Standard values.
    }
    \\
    \hfill
    \FramedSignature{\Change{\GG}}
    &
    \hfill
    \FramedSignature{\D \Gr \in \Change[\GG]{\Gr}}
    &
    \hfill
    \FramedSignature{\Gr \in \Eval{\GG}}
    \\
    \Align{
      \Change{\EmptyContext}
        & = \EmptyContext \\
      \Change{\Extend*{x}{\Gt}}
        & = \Extend[\Change{\GG}]{\D x}{\Change{\Gt}}
    }
    &
    \Align{
      \Change[\EmptyContext]{\EmptyEnv}
        & = \left\{ \EmptyEnv \right\} \\
      \Change[\Extend*{x}{\Gt}]{\ExtendEnv*{x}{v}}
        & = \left\{ \ExtendEnv*[\D \Gr]{\D x}{\D v} \mid \D \Gr \in \Change[\GG]{\Gr} \land \D v \in \Change[\Gt]{v} \right\}
    }
    &
    \Align{
      \Eval{\EmptyContext}
        & = \left\{ \EmptyEnv \right\} \\
      \Eval{\Extend{x}{\Gt}}
        & = \left\{ \ExtendEnv*{x}{v} \mid \Gr \in \Eval{\GG} \land v \in \Eval{\Gt}\right\}
    }
    \\
    \Subcaption{fig:correctness:change-contexts}{
      Change contexts.
    }
    &
    \Subcaption{fig:correctness:change-environments}{
      Change environments.
    }
    &
    \Subcaption{fig:correctness:environments}{
      Standard environments.\pg{Fix overfull hbox!}
    }
    \\
    \hfill
    \FramedSignature{\Change{t}}
    &
    \hfill
    \FramedSignature{\EvalIncSmashWith{t}{\Gr}{\D \Gr}}
    &
    \hfill
    \FramedSignature{\EvalWith{t}{\Gr}}
    \\
    \Align{
      \Derive{\Const{c}}
        & = \ldots\\
      \Derive{\Lam{x}{t}}
        & = \lambda x\;\D x.\ \Derive{t}\\
      \Derive{\App{s}{t}}
        & = \App{\App{\Derive{s}}{t}}{\Derive{t}}\\
      \Derive{\Var{x}}
        & = \Var{\D x}
    }
    &
    \Align{
      \EvalIncSmashWith{c}{\rho}{\D \rho}
        & = \ldots\\
      \EvalIncSmashWith{\Lam{x}{t}}{\rho}{\D \rho}
        & = \lambda v\;\D v.\ 
                   \EvalIncSmashWith
                   {t}
                   {\ExtendEnv*{x}{v}}
                   {\ExtendEnv*[\D \rho]{\D x}{\D v}}\\
      \EvalIncSmashWith{\App{s}{t}}{\rho}{\D \rho}
        & = \App
                   {\App
                     {\EvalIncSmashWith*{s}{\rho}{\D \rho}}
                     {\EvalWith*{t}{\rho}}}
                   {\EvalIncSmashWith*{t}{\rho}{\D \rho}}\\
      \EvalIncSmashWith{x}{\Gr}{\D \Gr}
        & = \textit{lookup $\D x$ in $\D \Gr$}
    }
    &
    \Align{
      \EvalWith{c}{\rho}
        & =\ldots\\
      \EvalWith{\Lam{x}{t}}{\rho}
        & = \lambda v.\ \EvalWith{t}{\ExtendEnv*{x}{v}}\\
      \EvalWith{\App{s}{t}}{\rho}
        & = \App{\EvalWith*{s}{\rho}}{\EvalWith*{t}{\rho}}\\
      \EvalWith{x}{\Gr}
        & = \textit{lookup $x$ in $\Gr$}
    }
    \\
    \Subcaption{fig:correctness:derive}{
      Differentiation.
    }
    &
    \Subcaption{fig:correctness:change-evaluation}{
      Differential evaluation.
    }
    &
    \Subcaption{fig:correctness:evaluation}{
      Standard evaluation.
    }
  \end{tabular}

  \caption{Standard and differential behavior of the simply-typed
    $\lambda$-calculus.
    The left column defines differentiation as a source-to-source
    transformation.
    The right column defines the standard semantics of the
    simply-typed lambda calculus.
    The middle column connects these artifacts via a differential
    semantics that maps $\Gl$-terms to the derivative of their
    standard semantics.}
  \label{fig:correctness}
\end{figure*}

\subsection{Architecture of the proof}

$\Derive{t}$ is defined using change types. As discussed in
\cref{ssec:change-types}, change types impose on their members
less restrictions than corresponding change structures -- they
contain ``junk'' (such as the change $-5$ for the natural number $3$). 
We cannot constrain the behavior of
$\Derive{t}$ on such junk; a direct correctness proof fails. To
avoid this problem, our proof defines a version of $\DERIVE$
which uses change structures instead.

To this end, we first present a standard denotational semantics
$\Eval{-}$ for simply-typed $\Gl$-calculus. Using our theory of
changes, we associate change structures to our domains. We define
a non-standard denotational semantics $\EvalInc{-}$, which is
analogous to $\DERIVE$ but operates on elements of change
structures, so that it needn't deal with junk. As a consequence,
we can prove that $\EvalInc{t}$ is the derivative of $\Eval{t}$:
this is our key result.

Finally, we define a correspondence between change sets and
domains associated with change types, and show that whenever
$\EvalInc{t}$ has a certain behavior on an input,
$\Eval{\Derive{t}}$ has the corresponding behavior on the
corresponding input. Our correctness property follows as a
corollary.

\subsection{Denotational semantics}
\label{sec:denotational-sem}

In order to prove that incrementalization preserves the meaning
of terms, we define a denotational semantics of the object
language. We first associate a domain with every type, given the
domains of base types provided by the plugin. Since our calculus
is strongly normalizing and all functions are total, we can
avoid using domain theory to model partiality: our domains are
simply sets. Likewise, we can use functions as the domain of function types.

\begin{definition}[Domains]
  The domain $\Eval{\Gt}$ of a type $\Gt$ is defined as in
  \cref{fig:correctness:values}.
\end{definition}

Given this domain
construction, we can now define an evaluation function for
terms. The plugin has to provide the evaluation function for
constants. In general, the evaluation function $\Eval{t}$ computes the value of a
well-typed term $t$ given the values of all free variables in
$t$. The values of the free variables are provided in an
environment.

\begin{definition}[Environments]
  An environment $\Gr$ assigns values to the names of free
  variables.

  \begin{syntax}
    \Gr ::= \EmptyContext \mid \ExtendEnv{x}{v}
  \end{syntax}

  We write $\Eval{\GG}$ for the set of environments that assign
  values to the names bound in $\GG$ (see
  \cref{fig:correctness:environments}).
\end{definition}

\begin{definition}[Evaluation]
  \label{def:evaluation}
  Given $\Typing{t}{\tau}$, the meaning of $t$ is defined by the
  function $\Eval{t}$ of type $\Fun{\Eval{\GG}}{\Eval{\tau}}$
  in \cref{fig:correctness:evaluation}.
\end{definition}

This is the standard semantics of the simply-typed
$\Gl$-calculus.
We can now specify what it means to incrementalize the
simply-typed $\Gl$ calculus with respect to this semantics.

\subsection{Change semantics}
The informal specification of differentiation is to map
changes in a program's input to changes in the program's
output. In order to formalize this specification in terms of
change structures and the denotational semantics of the object
language,
we now define a non-standard denotational semantics of the object
language that computes changes. The evaluation function
$\EvalInc{t}$ computes how the value of a well-typed term $t$
changes given both the values and the changes of all free
variables in $t$.
In the special case that none of the free variables change,
$\EvalInc{t}$ computes the nil change. By
\cref{thm:nil-is-derivative}, this is the derivative of
$\Eval{t}$ which maps changes to the input of $\Eval{t}$ to
changes of the output of $\Eval{t}$, as required.

First, we define a change structure on $\Eval{\Gt}$ for all
$\Gt$. The carrier $\Change[\Gt]$ of these change structures will
serve as non-standard domain for the change semantics. The plugin
provides a change structure $\ChangeStruct{C}_\Gi$ on base type
$\Gi$ such that $\forall v. \Change[\Gi]{v} \subseteq \Eval{\Change{\Gi}}$.

\begin{definition}[Changes]
  Given a type $\Gt$, we define a change structure
  $\ChangeStruct{C}_\Gt$ for $\Eval{\Gt}$ by induction on the
  structure of $\Gt$. If $\Gt$ is a base type $\Gi$, then
  the result $\ChangeStruct{C}_\Gi$ is supplied by the plugin.
  Otherwise we use the construction from \cref{thm:func-changestruct} and
  define
  \begin{align*}
    \ChangeStruct{C}_{\Fun{\Gs}{\Gt}} & = \ChangeStruct{C}_{\Gs} \to \ChangeStruct{C}_{\Gt}.
  \qedAligned
  \end{align*}
\end{definition}

To talk about the derivative of $\Eval{t}$, we need a change
structure on its domain, that is on the set of environments.
Since environments are (heterogeneous) lists of values, we
can lift operations on change structures to change structures on
environments, acting pointwise in the obvious way.

\begin{definition}[Change environments]
  \label{def:change-environments}
  Given a context $\GG$, we define a change
  structure $\ChangeStruct{C}_\GG$ on the corresponding
  environments $\Eval{\GG}$ and change environments $\Change[\GG]{\Gr}$
  in \cref{fig:correctness:change-environments}.

  The operations $\APPLY[\Gr]$ and $\DIFF[\Gr]$ are defined as follows.
  \begin{align*}
    \Apply{\EmptyContext}{\EmptyContext} &= {\EmptyContext} \\
    \Apply{\ExtendEnv*[\D \Gr]{\D x}{\D v}}{\ExtendEnv*{x}{v}} &= \ExtendEnv[\Apply*{\D \Gr}{\Gr}]{x}{\Apply*{\D v}{v}} \\[\eqsep]
    \Diff{\EmptyContext}{\EmptyContext} &= {\EmptyContext} \\
    \Diff{\ExtendEnv*[\Gr_2]{x}{v_2}}{\ExtendEnv*[\Gr_1]{x}{v_1}} &= \ExtendEnv[\Diff*{\Gr_2}{\Gr_1}]{x}{\Diff*{v_2}{v_1}}
  \end{align*}
  The properties in \Cref{def:change-struct} follow directly from the same properties
  for the underlying change structures $\ChangeStruct{C}_\Gt$.
\end{definition}

At this point, we can define the change semantics of terms and
prove that $\EvalInc{t}$ it is the derivative of $\Eval{t}$. For
each constant $c$, the plugin provides $\EvalInc{c}$, the derivative
of $\Eval{c}$.

\begin{definition}[Change semantics]
  \label{def:change-evaluation}
  The function $\EvalInc{t}$ is defined in
  \cref{fig:correctness:change-evaluation}.
\end{definition}

\begin{lemma}
  \label{lem:change-semantics-correct}
  Given $\Typing{t}{\Gt}$, $\EvalInc{t}$ is the derivative of $\Eval{t}$.
\end{lemma}

\begin{optionalproof}
\pg{This optional proof is phrased for fully applied constants.}
  Given a derivation of $\Typing{t}{\Gt}$, an environment $\Gr
  \in \Eval{\GG}$, and a corresponding change environments $\D
  \Gr \in \Change[\GG]{\Gr}$, we prove
  \[
    \Apply{\EvalIncWith*{t}{\Gr}{\D \Gr}}
          {\EvalWith*{t}{\Gr}}
    =
    \EvalWith{t}{\Apply*{\D \Gr}{\Gr}}
  \]
  by induction on the structure of the derivation of
  $\Typing{t}{\Gt}$. There is one case for each of the typing
  rules in \cref{def:typing}.

  \Case \Lam{x}{t_1}: In this case, $\Gt = \Fun{\Gs}{\Gt}$
  and we prove extensional equality of two functions. For an
  arbitrary value $v \in \Eval{\Gs}$, we have
  \begin{align*}
    &         \App
                {\Apply*
                  {\EvalIncWith{\Lam{x}{t_1}}{\Gr}{\D \Gr}}
                  {\EvalWith{\Lam{x}{t_1}}{\Gr}}}
                {v}\\
    & \quad = \Apply
                {\App
                  {\App
                    {\EvalIncWith*{\Lam{x}{t_1}}{\Gr}{\D \Gr}}
                    {v}}
                  {\Nil[\Gt _1]{v}}}
                {\App
                  {\EvalWith*{\Lam{x}{t_1}}{\Gr}}
                  {v}}\\
    & \quad = \Apply
                {\EvalIncWith
                  {t_1}
                  {\ExtendEnv*[\Gr]{x}{v}}
                  {\ExtendEnv*[\D \Gr]{\D x}{\Nil{v}}}}
                {\EvalWith
                  {t_1}
                  {\ExtendEnv*[\Gr]{x}{v}}}\\
    & \quad = \EvalWith
                {t_1}
                {\Apply*
                  {\ExtendEnv*[\D \Gr]{\D x}{\Nil{v}}}
                  {\ExtendEnv*[\Gr]{x}{v}}}\\
    & \quad = \EvalWith
                {t_1}
                {\ExtendEnv*
                  [\Apply{\D \Gr}{\Gr}]
                  {x}
                  {\Apply{\Nil{v}}{v}}}\\
    & \quad = \App
                {\EvalWith*
                  {\Lam{x}{t_1}}
                  {\Apply*{\D \Gr}{\Gr}}}
                {\Apply*{\Nil{v}}{v}}\\
    & \quad = \App
                {\EvalWith*
                  {\Lam{x}{t_1}}
                  {\Apply*{\D \Gr}{\Gr}}}
                {v}
  \end{align*}
  by
  \cref{def:function-changes:update,def:evaluation,def:change-evaluation},
  the induction hypothesis on $t_1$,
  \cref{def:change-environments,thm:update-nil}.

  \Case \App{s}{t_1}: We have
  \begin{align*}
    &         \Apply
                {\EvalIncWith{\App{s}{t_1}}{\Gr}{\D \Gr}}
                {\EvalWith{\App{s}{t_1}}{\Gr}}\\
    & \quad = \Apply
                {\App
                  {\App
                    {\EvalIncWith*{s}{\rho}{\D \rho}}
                    {\EvalWith*{t_1}{\rho}}}
                  {\EvalIncWith*{t_1}{\rho}{\D \rho}}}
                {\App
                  {\EvalWith*{s}{\Gr}}
                  {\EvalWith*{t_1}{\Gr}}}\\
    & \quad = \App
                {\Apply*
                  {\EvalIncWith{s}{\Gr}{\D \Gr}}
                  {\EvalWith{s}{\Gr}}}
                {\Apply*
                  {\EvalIncWith{t_1}{\Gr}{\D \Gr}}
                  {\EvalWith{t_1}{\Gr}}}\\
    & \quad = \App
                {\EvalWith*{s}{\Apply*{\D \Gr}{\Gr}}}
                {\EvalWith*{t_1}{\Apply*{\D \Gr}{\Gr}}}\\
    & \quad = \EvalWith
                {\App{s}{t_1}}
                {\Apply*{\D \Gr}{\Gr}}
  \end{align*}
  by
  \cref{def:evaluation,def:change-evaluation,thm:incrementalization}
  and the induction hypotheses on $s$ and $t_1$.

  \Case \Var{x}: Let $v$ be the entry for $\Var{x}$ in $\Gr$, and
  let $\D v$ be the entry for $\Var{\D x}$ in $\D \Gr$. We know that
  these entries exist from $\Typing{\Var{x}}{\Gt}$ with $\Gr \in
  \Eval{\GG}$ and $\D \Gr \in \Change[\GG]{\Gr}$. The entry for
  $\Var{x}$ in $\Apply{\D \Gr}{\Gr}$ is $\Apply{\D v}{v}$, so we
  have
  \begin{align*}
    &         \Apply
                {\EvalIncWith*{\Var{x}}{\Gr}{\D \Gr}}
                {\EvalWith*{\Var{x}}{\Gr}}\\
    & \quad = \Apply{\D v}{v}\\
    & \quad = \EvalWith{\Var{x}}{\Apply*{\Gr}{\D \Gr}}
  \end{align*}
  as required.

  \Case \Const{c}{\List{t}}: We have
  \begin{align*}
    &         \Apply
                {\EvalIncWith*{\Const{c}{\List{t}}}{\Gr}{\D \Gr}}
                {\EvalWith*{\Const{c}{\List{t}}}{\Gr}}\\
    & \quad = \Apply
                {\EvalIncConst*
                  {\Const{c}}
                  {\List*{\EvalWith{t}{\Gr}}}
                  {\List*{\EvalIncWith{t}{\Gr}{\D \Gr}}}}
                {\EvalConst*
                  {\Const{c}}
                  {\List*{\EvalWith{t}{\Gr}}}}\\
    & \quad = \EvalConst
                {c}
                {\List*{\EvalIncWith{t}{\Gr}{\D \Gr}}}\\
    & \quad = \EvalConst
                {c}
                {\List*{\EvalWith{t}{\Apply*{\D \Gr}{\Gr}}}}\\
    & \quad = \EvalWith
                {\Const{c}{\List{t}}}
                {\Apply*{\D \Gr}{\Gr}}
  \end{align*}
  by \cref{def:change-evaluation-const,def:change-evaluation} and
  the induction hypotheses on the terms $\List{t}$.
\end{optionalproof}

\subsection{Correctness of differentiation}
\label{sec:differentiate-correct}

\DeclareFixedFootnote{\EmptyEmptyNote}{%
To evaluate a closed term $t$, we need no environment entries, so
the empty environment $\EmptyEnv$ suffices:
$\EvalWith*{t}{\EmptyEnv}$ is the value of $t$ in the empty environment, and
$\;\EvalIncSmashWith*{t}{\EmptyEnv}{\EmptyEnv}$
is the value of $t$ using the change semantics, the empty environment and the empty change
environment.}

We can now
prove that the behavior of $\Eval{\Derive{t}}$ is consistent with
the behavior of $\EvalInc{t}$. This leads us to the proof of the
correctness theorem mentioned in the introduction.

The logical relation~\citep[Chapter 8]{Mitchell1996foundations}
of \emph{erasure} captures the idea that an
element of a change structure stays almost the same after we
erase all traces of dependent types from it.

\begin{definition}[Erasure]\label{def:erasure}
Let $\D v \in \Change[\Gt]v$ and $\D v' \in \Eval{\Change\Gt}$.
We say $\D v$ erases to $\D v'$, or
$\Implements[\Gt][v]{\D v}{\D v'}$,
if one of the following holds:
\begin{subdefinition}
\item $\Gt$ is a base type and $\D v=\D v'$.
\item $\Gt=\Fun{\Gs_0}{\Gs_1}$ and for all $w$, $\D w$, $\D w'$
such that $\Implements[\Gs_0][w]{\D w}{\D w'}$, we have
$\Implements[\Gs_1][\App* v w]
{\App*{\App{\D v}w}{\D w}}
{\App*{\App{\D v'}w}{\D w'}}$. \qed
\end{subdefinition}
\end{definition}

Sometimes we shall also say that $\D v \in \Change[\Gt] v$ erases
to a closed term $\HasType{\D t}{\GD t}$, in which case we mean
$\D v$ erases to $(\EvalWith{\D t}{\EmptyEnv})$.\EmptyEmptyNote

The following lemma makes precise what we meant by
``almost the same''.

\begin{lemma}\label{lem:almost-the-same}
Suppose $\Implements[\Gt][v]{\D v}{\D v'}$. If $\UPDATE'$ is the
erased version of the update operator $\UPDATE$ of the change
structure of $\Gt$ (\cref{ssec:change-types}), then
\[
v \UPDATE \D v = v \UPDATE' \D v'. \qed
\]
\end{lemma}

It turns out that $\EvalIncWith{t}$ and $\Derive{t}$ are ``almost the
same''. For closed terms, we make this precise by:

\begin{lemma}
  \label{lem:derive-correct}
If $(\HasType t \Gt)$ is closed, then $\EvalIncSmashWith*{t}\EmptyEnv\EmptyEnv$ erases to
$\Derive{t}$.
\end{lemma}

\begin{optionalproof}
  \pg{Say that we omit the proof because it's extremely boring.
    The recursion scheme seems less obvious than I'd have
    expected since it uses a logical relation but also deals with open terms.}
\end{optionalproof}

We omit for lack of space a more general version of
\cref{lem:derive-correct}, which holds also for open terms, but
requires defining erasure on environments.
The main correctness theorem is a corollary of
\cref{lem:almost-the-same,lem:derive-correct,lem:change-semantics-correct}.

\begin{theorem}[Correctness of differentiation]
\label{thm:main}
Let $\HasType{f}{\Fun \Gs \Gt}$ be a closed term of function
type. For every closed base term $\HasType{s}{\Gs}$ and for every closed change term
$\HasType{\D s}{\Change\Gs}$ such that some change
$\D v\in\Change[\Gs]{\Eval{s}}$ erases to $\D s$, we
have
\[
  \App{f}{\Update*{s}{\D s}}
\cong
  \Update{\App*{f}{s}}{\App*{\App{\Derive f}{s}}{\D s}},
\]
where $\cong$ is denotational equality ($a \cong b$ iff $\Eval{a} = \Eval{b}$).
\end{theorem}
\cref{thm:main} is a more precise restatement of
\cref{eq:correctness}. Requiring the existence of $\D v$ ensures
that $\D s$ evaluates to a change, and not to junk in
$\Eval{\Change\Gs}$.

\subsection{Plugins}\label{ssec:plugin}

\pg{Given enough time, we might want to move the plugin
  requirements here --- as long as we can keep the rest of the
  text consistent (for instance by explaining the role of ...).}
Both our correctness proof and the differentiation framework (which is the 
basis for our implementation) are parametric in the plugin. 
Instantiating the differentiation framework requires a \emph{differentiation plugin};
instantiating the correctness proof for it  requires a \emph{proof           plugin}.

\pg{Consider also describing (in one sentence) the language plugin (introducing constants and base types).}
To allow executing and differentiating $\Gl$-terms, a differentiation plugin must
provide:
\begin{itemize}
\item base types, and for each base type $\Gi$, the erased change structure of $\Gi$ as specified in
\cref{fig:change-operations},
\item primitives, and for each primitive $c$, the term $\Derive{c}$.
\end{itemize}

To instantiate the correctness proof to a plugin,
one must provide additional definitions and lemmas.
For each base type $\Gi$, a proof plugin must provide:
\begin{itemize}
\item a semantic domain $\Eval{\Gi}$,
\item a change structure $\ChangeStruct{C}_\Gi$ such that $\forall v. \Change[\Gi]{v} \subseteq \Eval{\Change{\Gi}}$,
\item a proof that $\ChangeStruct{C}_\Gi$ erases to the corresponding erased change structure in the differentiation plugin.
\end{itemize}
For each primitive $\HasType c \Gt$, the proof plugin must provide:
\begin{itemize}
\item its value $\Eval{c}$ in the domain $\Eval{\Gt}$,

\item its derivative $\EvalIncSmashWith*{c}{\EmptyEnv}{\EmptyEnv}$\EmptyEmptyNote{} in the change set of $\Gt$,
\item a proof that $\EvalIncSmashWith*{c}{\EmptyEnv}{\EmptyEnv}$ erases to the term $\Derive{c}$.
\end{itemize}

To show that the interface for proof plugins
can be implemented, we wrote a small proof plugin with
integers and bags of integers\yc{link to agda}.
To show that differentiation plugins are practicable, we 
have implemented the transformation and a differentiation plugin
which allows the incrementalization of non-trivial programs.
This is presented in the next section.

\section{Differentiation in practice}
\label{sec:applying}

In practice, successful incrementalization requires both
correctness and performance of the derivatives. Correctness of
derivatives is guaranteed by the theoretical development the
previous sections, together with the interface for
differentiation and proof plugins, whereas performance of
derivatives has to come from careful design and implementation of
differentiation plugins.

\subsection{The role of differentiation plugins}
\label{ssec:methodology}
Users of our approach need to
(1) choose which base types and primitives they need,
(2) implement suitable differentiation plugins for these base
types and primitives,
(3) rewrite (relevant parts of) their programs in terms of these primitives and
(4) arrange for their program to be called on changes instead of
updated inputs.

As discussed in \cref{ssec:differentiation}, differentiation
supports abstraction, application and variables, but since
computation on base types is performed by primitives for those
types, efficient derivatives for primitives are essential for
good performance.

To make such derivatives efficient, change
types must also have efficient implementations, and allow
describing precisely what changed. The efficient derivative of
$\Term{sum}$ in \cref{sec:intro} is possible only if bag changes
can describe deletions and insertions, and integer changes can
describe additive differences.

For many conceivable base types, we do not have to design the
differentiation plugins from scratch. Instead, we can reuse the
large body of existing research on incrementalization in
first-order and domain-specific settings. For instance, we reuse
the approach from \citet{GlucheGrust97Incr} to support incremental
bags and maps. By wrapping a
domain-specific incrementalization result in a differentiation
plugin, we adapt it to be usable in the context of a higher-order
and general-purpose programming language, and in interaction with
other differentiation plugins for the other base types of that
language.

\begin{figure*}
\includegraphics{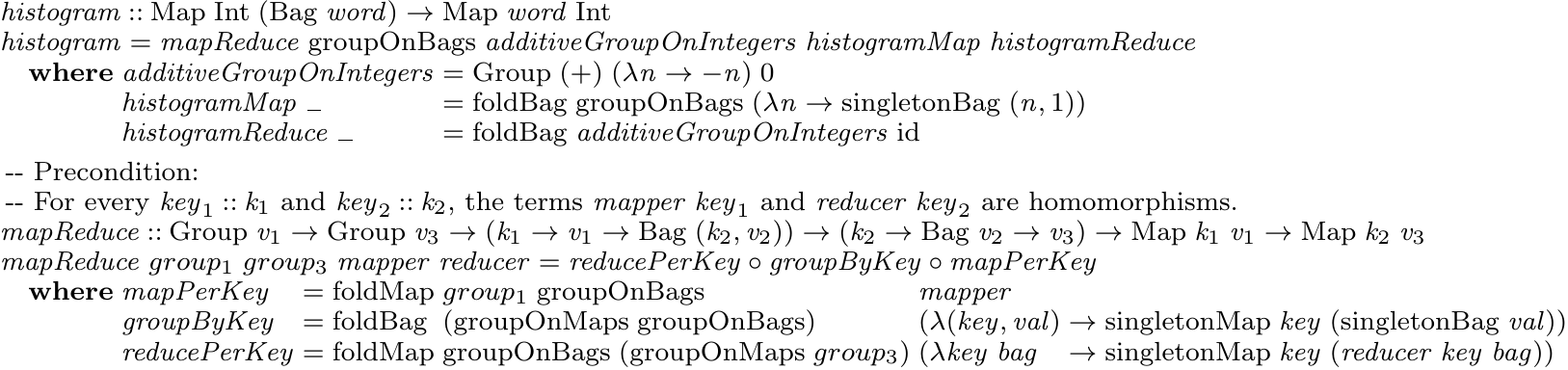}
\caption{The $\Gl$-term $\HISTOGRAM$ with Haskell-like syntactic sugar.
$\Term{additiveGroupOnIntegers}$ is the group on integers described
in \cref{ssec:change-structures}.}
\label{fig:case-study-pseudocode}
\end{figure*}

For base types with no known incrementalization strategy,
the precise interfaces for differentiation and
proof plugins can guide the implementation effort. These
interfaces could also from the basis for a library of
differentiation plugins that work well together.

Rewriting whole programs in our language would be an excessive
requirements. Instead, we embed our object language as an EDSL in
some more expressive meta-language (Scala in our case study), so
that embedded programs are reified. The embedded language can be
made to resemble the metalanguage~\citep{rompf2010lightweight}.
To incrementalize a part of a computation, we write it in our
embedded object language, invoke $\DERIVE$ on the embedded
program, optionally optimize the resulting programs and finally
invoke them. The metalanguage also acts as a macro system for the
object language, as usual. This allows us to simulate polymorphic
collections such as $\Bag*[\Gi]$ even though the object language is
simply-typed; technically, our plugin exposes a family of base
types to the object language.

\pg{Explain how we construct changes? The metalanguage can then
  construct changes in different ways?}

\subsection{Predicting nil changes}
Handling changes to all inputs can induce excessive overhead in incremental
programs\citep{Acar09}. It is also often
unnecessary; for instance, the function argument of $\Term{fold}$ in 
\cref{sec:intro} does not change since it is a closed subterm of the program, so
$\Term{fold}$ will receive a nil change for it.
A (conservative) static analysis can detect changes that are guaranteed to be nil at runtime. 
We can then specialize derivatives that receive this change, so that they need not
inspect the change at runtime.

For our case study, we have implemented a simple static analysis which
detects and propagates information about closed terms. The analysis is not interesting and 
we omit details for lack of space.

\subsection{Self-maintainability}
\label{sec:performance-cons}
\label{ssec:self-maint}

In databases, a self-maintainable view~\citep{Gupta99MMV} is a function that can
update its result from input changes alone, without looking at
the actual input. By analogy, we call a derivative
\emph{self-maintainable} if it uses no
base parameters, only their changes. Self-maintainable derivatives
describe efficient incremental computations: since they do not
use their base input, their running time does not have to depend on the input
size.

\pg{Dependency: bags, their changes and primitives.}

For instance, $\UNION$ on bags is self-maintainable with the
change structure $\ChangeStruct{\Bag{\Gi}}$ described in
\cref{ssec:change-structures}; its derivative
$\Derive{\UNION} = \Lam* {x\; \D x\; y \; \D y}
      {\Union{\D x}{\D y}}$ does not use the base
inputs $x$ and $y$.
On the other hand, $\FOLDBAG$ is not necessarily self-maintainable.
However, $\App* \FOLDBAG f$ is self-maintainable if we can predict
that changes to $f$ are going to be nil\pg{Elaborate; show the
  derivative, ensure prerequisites are there}.
We take advantage of this by implementing a specialized
derivative.

To avoid recomputing base arguments for self-maintainable derivatives
(which never need them), we
currently employ lazy evaluation.  Since we could use standard techniques for dead-code
elimination~\citep{Appel97} instead, laziness is not central to our
approach, however.
Derivatives which are not self-maintainable  need their base
arguments, which can be expensive to compute. Since they are also
computed while running the base program, one could reuse the previously
computed value through memoization or extensions of static
caching (as discussed in \cref{sec:rw}). We leave implementing these optimizations for future work. As a consequence,
our current implementation delivers good results only if
most derivatives are self-maintainable.

\subsection{Case study}
\label{sec:plugins}

To demonstrate that \ILC\ can speed up realistic
programs, we perform a case study on a nontrivial one.
We take the MapReduce-based skeleton of the
word-count example, as described by \citet{Lammel07}. We define a
suitable differentiation plugin, adapt the program to use it and show
that incremental computation is faster than recomputation on it.
We designed and implemented the differentiation plugin 
following the requirements on the corresponding proof plugin,
even though we did not yet formally (for example, in Agda) define
the proof plugin. For lack of space, we focus on base types
which are crucial for our example and its performance, that is,
collections.
The plugin also implements tuples, tagged unions, Booleans and
integers with the usual introduction and elimination forms, with
few optimizations for their derivatives.

$\WORDCOUNT$ takes a map
from document IDs to documents and produces a map
from words appearing in the input to the count of their
appearances, that is, a histogram:
\[
\HasType \WORDCOUNT {\Fun {\HashMap \DOCUMENTID \DOCUMENT} {\HashMap \WORD \Int}}
\]
For simplicity, instead of modeling strings, we model documents
as bags of words and document IDs as integers. Hence, what we implement is:
\[
\HasType \HISTOGRAM {\Fun {\HashMap \Int {\Bag*[a]}} {\HashMap a \Int}}
\]
We model words by integers ($a = Int$), but treat them parametrically.
Other than that, we adapt directly
\citeauthor{Lammel07}'s code to our language.
\pg{But it cannot accept all the same parameters... Ah, but it
  depends on which foldMap we call. OK.}
\Cref{fig:case-study-pseudocode} shows the $\Gl$-term
$\HISTOGRAM$.

\begin{figure}

\centering

\begin{lstlisting}[language=scala,
firstline=10,
escapechar=|,
literate=
{=>}{$\Rightarrow\,$}2
{>=}{$\ge\;$}2
{<-}{$\leftarrow\;$}2
{!=}{$\ne\;$}2
{Abelian-group-based}{\text{Abelian-group-based }}1
{<}{$<$}2
{bag1}{{bag$_1$}}4
{bag2}{{bag$_2$}}4
{value1}{{value$_1$}}6
{value2}{{value$_2$}}6
{dict1}{{dict$_1$}}5
{dict2}{{dict$_2$}}5
{v1}{{v$_1$}}2
{v2}{{v$_2$}}2
{???}{$\ldots$}3
]
// Abelian groups
abstract class Group[A] {
  def merge(value1: A, value2: A): A
  def inverse(value: A): A
  def zero: A
}

// Bags
type Bag[A] = collection.immutable.HashMap[A, Int]

def groupOnBags[A] = new Group[Bag[A]] {
  def merge(bag1: Bag[A], bag2: Bag[A]) = ???
  def inverse(bag: Bag[A]) = bag map {
    case (value, count) => (value, -count)
  }
  def zero = collection.immutable.HashMap()
}

def foldBag[A, B](group: Group[B], f: A => B, bag: Bag[A]): B =
  bag.flatMap {
    case (x, c) if c >= 0 => Seq.fill(c)(f(x))
    case (x, c) if c < 0 => Seq.fill(-c)(group.inverse(f(x)))
  }.fold(group.zero)(group.merge)

// Maps
type Map[K, A] = collection.immutable.HashMap[K, A]

def groupOnMaps[K, A](group: Group[A]) = new Group[Map[K, A]] {
  def merge(dict1: Map[K, A], dict2: Map[K, A]): Map[K, A] =
    (dict1 merged dict2) {
      case ((k, v1), (_, v2)) => (k, group.merge(v1, v2))
    } filter {
      case (k, v) => v != group.zero
    }

  def inverse(dict: Map[K, A]): Map[K, A] = dict map {
    case (k, v) => (k, group.inverse(v))
  }

  def zero = collection.immutable.HashMap()
}

// The general map fold
def foldMapGen[K, A, B](zero: B, merge: (B, B) => B)
  (f: (K, A) => B, dict: Map[K, A]): B =
    dict.map(Function.tupled(f)).fold(zero)(merge)

// By using |\texttt{foldMap}| instead of |\texttt{foldMapGen}|, the user promises that
// |\texttt{f k}| is a homomorphism from |\texttt{groupA}| to |\texttt{groupB}| for each |\texttt{k : K}|.
def foldMap[K, A, B](groupA: Group[A], groupB: Group[B])
  (f: (K, A) => B, dict: Map[K, A]): B =
    foldMapGen(groupB.zero, groupB.merge)(f, dict)
\end{lstlisting}
\caption{A Scala implementation of primitives for bags and maps.
In the code, we call $\boxplus$, $\boxminus$ and $e$ respectively \emph{merge}, \emph{inverse}, and \emph{zero}.
We also omit the relatively standard primitives.}
\label{fig:primitives}
\end{figure}

\pg{Why does 'target language' show up all of a sudden? Shouldn't
  we say that we generate Scala code for execution
  \textbf{elsewhere}? Right now, we say it in next section. Confusing.}
\Cref{fig:primitives} shows a simplified
Scala implementation of the primitives
used in \cref{fig:case-study-pseudocode}.%
\pg{Klaus asked that I rewrite the text for understandability. I got started.}
As bag primitives, we provide constructors and a fold operation,
following \citet{GlucheGrust97Incr}. The constructors for bags
are $\Empty$ (constructing the empty bag), $\SINGLETON$
(constructing a bag with one element), $\UNION$ (constructing the
union of two bags) and $\NEGATE$ ($\Negate{b}$ constructs a bag
with the same elements as $b$ but negated multiplicities); all but $\SINGLETON$ represent
abelian group operations.
Unlike for usual ADT constructors, the same bag can be
constructed in different ways, which are equivalent by the equations defining abelian groups;
for instance, since $\UNION$ is commutative, $\Union{x}{y} = \Union{y}{x}$.
Folding on a bag will represent the bag through constructors in
an arbitrary way, and then replace constructors with arguments;
to ensure a well-defined result, the arguments of fold should
respect the same equations, that is, they should form an abelian group;
for instance, the binary operator should be commutative.
Hence, the fold operator $\FOLDBAG$ can be defined to take a
function (corresponding to $\SINGLETON$) and an abelian group
(for the other constructors). $\FOLDBAG$ is then defined by equations:
\begin{alignat*}{2}
&  \HasType{\FOLDBAG}{\mathbf{Group}\; \Gt \to (\Gs \to \Gt) \to &&\Bag[\Gs] \to \Gt}\\
&  \App{\FoldBag {g @ (\_, \boxplus, \boxminus, e)}{f}}{\Empty}
    &&= e \displaybreak[0]\\
&  \App{\FoldBag {g @ (\_, \boxplus, \boxminus, e)}{f}}{\Union* {b_1} {b_2}}
   &&= \App{\FoldBag{g}{f}}{b_1} \displaybreak[0]\\
&&& \boxplus
  \App{\FoldBag{g}{f}}{b_1} \displaybreak[0]\\
&  \App{\FoldBag {g @ (\_, \boxplus, \boxminus, e)}{f}}{\Negate* b}
   && = \boxminus \; \App*{\FoldBag{g}{f}}{b}\displaybreak[0]\\
&  \App{\FoldBag {g @ (\_, \boxplus, \boxminus, e)}{f}}{\Singleton*{v}}
   &&= \App{f}{v}
\end{alignat*}
If $g$ is a group,
these equations specify $\App{\FOLDBAG} g$ precisely~\citep{GlucheGrust97Incr}.
Moreover, since $\FoldBag{g}{f}$ satisfies the first three
equations, it satisfies the definition of an \emph{abelian group
  homomorphism} between the abelian group on bags and the group
$g$ (because those equations coincide with the definition).
\Cref{fig:primitives} shows an implementation of $\FOLDBAG$ as
specified above.
Moreover, all functions which deconstruct a bag can be expressed in
terms of $\FOLDBAG$ with suitable arguments.
For instance, we can sum the elements of a bag of integers with
$\FoldBag{\Term{gZ}}{\Lam*{x}{x}}$, where
$\Term{gZ}$ is the abelian group on integers defined in
\cref{ssec:change-structures}.
Users of $\FOLDBAG$ can define different abelian groups to specify
different operations (for instance, to multiply floating-point numbers).

If $g$ and $f$ do not change, $\FoldBag{g}{f}$ has a self-maintainable
derivative.
By the equations above,
\pg{Non-standard alignment, because the last line doesn't fit.}
\begin{align*}
& \FoldBag{g}{f} \Update*{b}{\D b}\\
=\;& \FoldBag{g}{f} {(\Union b {\D b})}\displaybreak[0]\\
=\;& \App{\FoldBag{g}{f}}{b}  \boxplus \App{\FoldBag{g}{f}}{\D b} \displaybreak[0]\\
=\;& \Update{\App{\FoldBag{g}{f}}{b}}{\Term{GroupChange} \; g \; \left(\App{\FoldBag{g}{f}}{\D b}\right)}
\end{align*}
We will describe the $\Term{GroupChange}$ change constructor in a moment.
Before that, we note that as a consequence, the derivative of $\FoldBag{g}{f} $ is
\[
\Lam{b \; db} \Term{GroupChange} \; g \; \left(\App{\FoldBag{g}{f}}{\D b}\right)\text{,}
\]
and we can see it does not use $b$: as desired, it is
\emph{self-maintainable}. Additional restrictions are require to
make $\FOLDMAP$'s derivative self-maintainable. Those restrictions require the
precondition on $\Term{mapReduce}$ in
\cref{fig:case-study-pseudocode}. $\Term{foldMapGen}$ has the
same implementation but without those restrictions; as a
consequence, its derivative is not self-maintainable, but it is more generally applicable.
Lack of space prevents us from giving more details.

To define $\Term{GroupChange}$, we need a suitable erased change
structure on $\Gt$, such that $\UPDATE$ will be equivalent to
$\boxplus$. Since there might be multiple groups on $\Gt$, we
\emph{allow the changes to specify a group}, and have
$\UPDATE$ delegate to $\boxplus$:
\begin{align*}
& \Change{\Gt} = \Term{Replace}\; \Gt \mid \Term{GroupChange} \Abelian*{\Gt} \Gt \\
& \Update{v}{(\Term{Replace} \; u)} = u\\
& \Update{v}{(\Term{GroupChange} \; (\bullet, \Term{inverse}, \Term{zero})\; dv)} = v \bullet dv\\
& \Diff{v}{u} = \Term{Replace}\; v
\end{align*}
That is, a change between two values is either simply the new
value (which replaces the old one, triggering recomputation),
or their difference (computed with abelian group
operations, like in the changes structures for groups from
\cref{ssec:change-structures}. The operator $\DIFF$
does not know which group to use, so it does not take advantage
of the group structure.
However, $\FOLDBAG$ is now able to generate a group change.
\pg{Clarify about user-defined groups.}

\subsection{Benchmarks}

Benchmarking our case study shows
that \ILC\ can offer order-of-magnitude speedups for a
realistic higher-order program.

\paragraph{Benchmarking setup}
\pg{Omit this at least while we lack space.}

We run object language programs by generating corresponding Scala code.
To ensure rigorous
benchmarking~\citep{Georges07rigorousJavaPerformance}, we use the
Scalameter benchmarking library. To show that the performance
difference from the baseline is statistically significant, we
show 99\%-confidence intervals in graphs.

We verify \cref{eq:correctness} experimentally
by checking that the two sides of the equation always
evaluate to the same value.

\paragraph{Input generation}

Inputs are randomly generated to resemble English words over all
webpages on the internet: The vocabulary size and the average
length of a webpage stay relatively the same, while the number of
webpages grows day by day. To generate a size-$n$ input of type
$(\HashMap{\Int}{\Bag*[\Int]})$, we generate $n$ random numbers
between 1 and 1000 and distribute them randomly in $n/1000$ bags.
Changes are randomly generated to resemble edits. A change has
50\% probability to delete a random existing number, and has 50\%
probability to insert a random number at a random location.
\pg{Is this the same $n$ as in the graph? I'm guessing the graphs
  show $n/1000$ or something like that. Cai, let's revise this
  together.}

\paragraph{Experimental units}

Thanks to \cref{eq:correctness}, both recomputation
$\App{f}{\Apply*{\D a}{a}}$ and incremental computation
$\Apply{\App*{\App{\Derive{f}}{a}}{\D a}}{\App*{f}{a}}$ produce
the same result. To show that derivatives are faster, we compare
these two computations. To compare with recomputation, we measure the
\emph{aggregated} running time for running the derivative on the change 
and for updating the original output with the result of the derivative.

\pg{Once we explain what the
  program does, refine this section describing the input
  generation, the changes in more detail, and so on.}

\subsection{Experimental results}

\pg{Warning: benchmark results are hardcoded here, copy-pasted
  from the spreadsheet. Edit with care.}

\begin{figure}
\includegraphics[keepaspectratio,width=8.5cm]{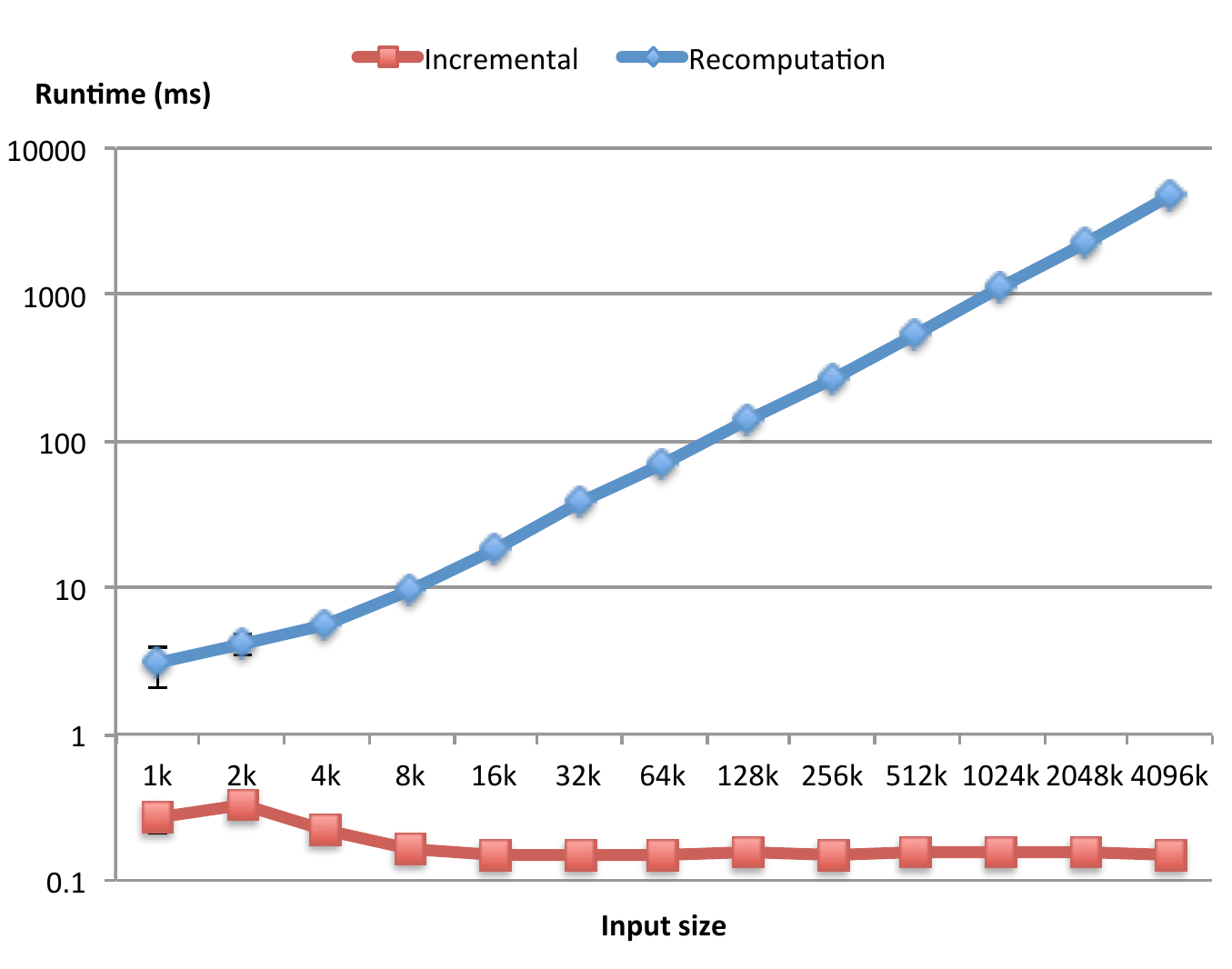}
\caption{Performance results in log-log scale, with input size on
  the x-axis and runtime in ms on the y-axis. Confidence
  intervals are shown by the whiskers; most whiskers are simply
  too small to be visible.}
\label{fig:graph}
\end{figure}

We present our results in \cref{fig:graph}. As expected, the
runtime of incremental computation is
\emph{essentially constant} in the size of the input, while the runtime
of recomputation is linear in the input size.
Hence, on our biggest inputs\pg{specify} incremental computation
is over $10^4$ times faster.

Derivative time is in fact slightly irregular for the first few inputs,
but this irregularity decreases slowly with increasing warmup
cycles. In the end, for derivatives we use $10^4$ warmup cycles.
With fewer warmup cycles, running time for derivatives decreases
significantly during execution, going from 2.6ms for $n = 1000$ to
0.2ms for $n = 512000$. Hence, we believe extended warmup is
appropriate, and the changes do not affect our general
conclusions. Considering confidence intervals, in our experiments the running
time for derivatives varies between 0.139ms and 0.378ms.

In our current implementation, the code of the generated derivatives can become 
quite big. For the histogram example (which is around 1KB of code), a pretty-print
of its derivative is around 40KB of code. The function application case
in \cref{fig:correctness:derive} can lead to a quadratic growth in the 
worst case. We believe that the code size of the derivative can be reduced again 
by common subexpression elimination, but we did not yet pursue that option.

\paragraph{Summary}
Our results show that the incrementalized program runs
in essentially constant time and hence orders of magnitude faster than
the alternative of recomputation from scratch. 

\section{Related work}
\label{sec:rw}

Existing work on incremental computation can be divided into two
groups: Static incrementalization and dynamic incrementalization.
Static approaches analyze a program statically and generate an incremental
version of it. Dynamic approaches create dynamic dependency graphs while
the program runs and propagate changes along these graphs.

The trade-off between the two is that static approaches have the potential
to be faster because no dependency tracking at runtime is needed, whereas
dynamic approaches can support more powerful programming languages.
The quick summary of how \ILC\ fits into this landscape is that
it pushes the envelope with regard to the expressive power of languages
whose programs can be incrementalized statically.

In the remainder of this section, we analyze the relation to the
most closely related prior works. \Citet{Ramalingam93} and
\citet{Acar06} discuss further related work.

\subsection{Dynamic approaches}
One of the most advanced dynamic approach to incrementalization is
self-adjusting computation, which has been applied to Standard ML
and large subsets of C~\citep{Acar09,Hammer11}.
In this approach, programs execute on the original
input in an enhanced runtime environment that tracks the
dependencies between values in a \emph{dynamic
  dependence graph}~\citep{Acar06}; intermediate results are
memoized.
Later, changes to the input propagate through
dependency graphs from changed inputs to results,
updating both intermediate and final results;
this processing is often more efficient than recomputation.

However, creating dynamic
dependence graphs imposes a large constant-factor overhead during
runtime, ranging from 2 to 30 in reported
experiments~\citep{Acar09EAS,Acar10TDT}, and affecting the
initial run of the program on its base input.
\citet{Acar10TDT} show how to support high-level data
types in the context of self-adjusting computation; however, the
approach still requires expensive runtime bookkeeping during the initial run.
Like other static approaches, our work needs no modified runtime
environment and has no overhead
during base computation, though it may be less efficient when processing
changes. This pays off if the initial input is 
big compared to its changes.

\citet{Chen11} have developed a static transformation for purely
functional programs, but this transformation just provides a superior inferface to use
the runtime support with less boilerplate, and does not reduce
this performance overhead. Hence, it is still a dynamic approach and 
should not be confused with the transformation we show in this work.

Another property of self-adjusting computation
is that incrementalization is only efficient if the program has a suitable
computation structure. For instance, a program
folding the elements of a bag with a left or right fold will not
have efficient incremental behavior; instead, it's necessary that
the fold be shaped like a balanced tree. In general,
incremental computations become efficient only if they are \emph{stable}~\citep{Acar05}.
Hence one may need to massage the program to make it efficient. Our methodology is 
different: Since we do not aim to incrementalize arbitrary programs written in standard
programming languages, we can select primitives that have efficient derivatives and thereby require 
the programmer to use them.

Functional reactive programming \citep{Elliott:1997:FRA:258948.258973}
can also be seen as a dynamic approach to incremental computation;
recent work by \citet{Maier2013} has
focused on speeding up reactions to input changes by making them
incremental on collections. Dynamic techniques are also used by
\citet{Willis08} to incrementalize JQL queries.

\subsection{Static approaches}
\pg{If we discuss partial evaluation, we should compare to
  \citep{Sundaresh91}}
Static approaches analyze a program at compile-time and produce an
incremental version that efficiently updates the output
of the original program according to changing inputs.

Static approaches have the potential to be more efficient than dynamic approaches,
because no bookkeeping at runtime is required. Also, the computed incremental
versions can often be optimized using standard compiler techniques
such as constant folding or inlining.
However, none of them support first-class functions; some
approaches have further restrictions.

Our aim is to apply static incrementalization to more expressive languages;
in particular, \ILC\ supports first-class functions and an open
set of base types with associated primitive operations.

\subsubsection{Finite differencing}
\label{sec:finite-diff}
Our work and terminology is partially inspired by \emph{finite differencing}~\citep{Paige82FDC}. 
\citet{Paige82FDC} present derivatives for a first-order language with a 
fixed set of primitives. This work has inspired variants of finite differencing
for queries on relational data, such as \emph{algebraic differencing}~\citep{Gupta99MMV}, and
\emph{delta processing}~\citep{Koch10IQE}.

However, most work in the database community is specialized to
relational databases, hence does not support nested data (either
nested collections, or algebraic data types). Incremental support
is further designed monolithically for a whole language, rather
than piecewise. The languages that are considered do not support
first-class functions.

More general (non-relational) data types are considered in the work by \citet{GlucheGrust97Incr};
our support for bags and the use of groups is inspired by their work,
but their architecture is still rather restrictive: they lack
support for function changes and restrict incrementalization to
self-maintainable views.

\pg{Think about adding what follows:
As we will see in Sec.~\ref{sec:differentiate}, adding support
for first-class functions means that what we add to this work
is better support for composing primitives together.}

\subsubsection{Static memoization}
\citet{Liu00}'s work allows to incrementalize a first-order base
program $f(\Old{x})$ to compute $f(\New{x})$, knowing how
$\New{x}$ is related to $\Old{x}$. To this end, they transform
$f(\New{x})$ into an incremental program which reuses the
intermediate results produced while computing $f(\Old{x})$, the
base program. To this end, (i) first the base program is
transformed to save all its intermediate results, then (ii) the
incremental program is transformed to reuse those intermediate
results, and finally (iii) intermediate results which are not
needed are pruned from the base program. However, to reuse
intermediate results, the incremental program must often be
rearranged, using some form of equational reasoning, into some
equivalent program where partial results appear literally. For
instance, if the base program $f$ uses a left fold to sum the
elements of a list of integers $\Old{x}$, accessing them from the
head onwards, and $\New{x}$ prepends a new element $h$ to the
list, at no point does $f(\New{x})$ recompute the same results.
But since addition is commutative on integers, we can rewrite
$f(\New{x})$ as $f(\Old{x}) + h$. The author's CACHET system will
try to perform such rewritings automatically, but it is not
guaranteed to succeed. Similarly, CACHET will try to synthesize
any additional results which can be computed cheaply by the base
program to help make the incremental program more efficient.

Since it is hard to fully automate such reasoning, we move
equational reasoning to the plugin design phase. A 
plugin provides general-purpose higher-order primitives for which
the plugin authors have devised efficient derivatives (by using
equational reasoning in the design phase). Then, the
differentiation algorithm computes incremental
versions of user programs without requiring further user intervention.
It would be useful to combine \ILC\ with some form of static
caching to make the computation of derivatives which
are not self-maintainable more efficient. We plan to do so
in future work.

\section{Conclusions and future work}
\label{sec:concl}
\label{ssec:future}
We have presented \ILC, an approach to lifting incremental computations
on first-order programs to incremental computations on higher-order
programs. We have presented a machine-checked correctness proof 
of a formalization of \ILC\ and an initial experimental evaluation
in the form of an implementation, a sample plugin for maps and bags,
and a non-trivial example that was incrementalized successfully and
efficiently. 

Our work opens several avenues of future work. Our current implementation
is not very efficient on derivatives that are not self-maintainable.
However, as discussed
(Sec.~\ref{ssec:self-maint}), we plan to investigate approaches
to memoizing intermediate results to address this limitation. Our next
step will be to develop language plugins which
have efficient non-self-maintainable primitives.

Another area of future work is adding support for algebraic data
types (including recursive types), polymorphism, subtyping, general recursion
and other collection types. While support for algebraic data
types could subsume support for specific collections, many
collections have additional algebraic properties that enable faster
incrementalization (like bags). Even lists (which have fewer algebraic properties)
can benefit from special support~\citep{Maier2013}.

Finally, we intend to perform a full and thorough experimental evaluation
to demonstrate that \ILC\ can incrementalize large-scale practical programs.

\iftoggle{names}{
\acks
We would like to thank Sebastian Erdweg, Ingo Maier, Erik Ernst, Tiark Rompf, and other ECOOP 2013
participants for helpful discussions.
This work is supported by the European Research Council, grant \#203099 ``ScalPL''.

}{}

\bibliographystyle{abbrvnat}

\bibliography{../../Bibs/ProgLang,../../Bibs/DB,../../Bibs/own,../../Bibs/SoftEng}

\appendix

\end{document}